\documentclass[useAMS,usenatbib]{mn2e}
\usepackage{times}
\usepackage{graphicx}
\usepackage{amssymb}

\newcommand{\msun}{$M_\odot$}
\newcommand{\lsun}{$L_\odot$}
\newcommand{\mbh}{$M_{\rm BH}$}
\newcommand{\sig}{$\sigma$}
\newcommand{\msig}{$M_{\rm BH}$-$\sigma$}
\newcommand{\mlum}{${M_{\rm BH}}$-$L$}
\newcommand{\degree}{$^{\circ}$}
\newcommand{\mlb}{$\Upsilon_b$}
\newcommand{\mld}{$\Upsilon_d$}
\newcommand{\ml}{$\Upsilon$}
\newcommand{\mlssp}{$\Upsilon_{\rm SSP}$}
\newcommand{\dchi}{$\Delta\chi^2$}

\title[The central black hole mass of NGC 1332]{The central black hole mass of the high-$\sigma$ but low-bulge-luminosity lenticular galaxy NGC 1332\thanks{Based on observations at the European Southern Observatory Very Large Telescope [082.B-0037(A)]}}

\author[S. P. Rusli et al.]{S.P.~Rusli,$^{1,2}$ J.~Thomas,$^{1,2}$ P.~Erwin,$^{1,2}$ R.P.~Saglia,$^{1,2}$ N.~Nowak,$^{1,2}$ R.~Bender$^{1,2}$  \\
$^1$ Max-Planck-Insitut f\"{u}r extraterrestrische Physik, Giessenbachstrasse, 85741 Garching, Germany\\
$^2$ Universit\"{a}tssternwarte, Scheinerstrasse 1, 81679 Munich, Germany}

\begin{document}


\maketitle

\label{firstpage}

\begin{abstract}
The masses of the most massive supermassive black holes (SMBHs) predicted by the \msig\ and \mlum\ relations appear to be in conflict. Which of the two relations is the more fundamental one remains an open question. NGC 1332 is an excellent example that represents the regime of conflict. It is a massive lenticular galaxy which has a bulge with a high velocity dispersion \sig\ of  $\sim320$ km s$^{-1}$; bulge--disc decomposition suggests that only 44\% of the total light comes from the bulge. The \msig\ and the \mlum\ predictions for the central black hole mass of NGC 1332 differ by almost an order of magnitude. We present a stellar dynamical measurement of the SMBH mass using an axisymmetric orbit superposition method. Our SINFONI integral-field unit (IFU) observations of NGC 1332 resolve the SMBH's sphere of influence which has a diameter of $\sim$0.76 arcsec. The \sig\ inside 0.2 arcsec reaches $\sim400$ km s$^{-1}$. The IFU data allow us to increase the statistical significance of our results by modelling each of the four quadrants separately. We measure a SMBH mass of ($1.45\pm0.20)\times10^9$\msun\ with a bulge mass-to-light ratio of $7.08\pm0.39$ in the $R$-band. With this mass, the SMBH of NGC 1332 is offset from the \mlum\ relation by a full order of magnitude but is consistent with the \msig\ relation. 
\end{abstract}


\section[]{Introduction}
It is now widely accepted that massive elliptical galaxies and (classical) bulges of spiral galaxies harbour supermassive black holes (SMBHs) at their centres with masses ranging between $\sim10^6$-$10^{10}$\msun. An increasing number of SMBH detections have led to the discovery of empirical correlations between the SMBH mass \mbh\ and the velocity dispersion \sig\ or the luminosity $L$ of the host bulge (\citealt{Kormendy-95}; \citealt{Magorrian-98}; \citealt{Ferrarese-00}; \citealt{Gebhardtletter-00}; \citealt{Tremaine-02}; \citealt{Marconi-03}; \citealt{Haering-04}; \citealt{Gueltekin-09}). These correlations indicate a strong connection between the SMBH and the host galaxy. Consequently, SMBH studies are essential for a better understanding of galaxy formation processes \citep{Silk-98}.

Because of its crucial role, SMBH demographics has become one of the key ingredients in cosmological simulations and theoretical models in recent years (\citealt{Granato-04}; \citealt{Somerville-08}; \citealt{Dimatteo-05}). An inventory of the SMBH population is therefore necessary. The number of secure measurements is presently only $\sim50$ and quiescent or weakly active SMBH beyond the local universe can typically not be well-studied. In this situation, the \msig\ and \mlum\ relations have become valuable tools in predicting the SMBH mass and mass function. By combining the distribution of the readily observed parameter $\sigma$ (or $L$) with the corresponding \msig\ (or \mlum) relation, the SMBH abundance can be indirectly estimated.

This method, however, suffers from a bias: the \msig\ relation predicts fewer SMBHs with masses $\gtrsim 10^9M_\odot$ than does the \mlum\ relation. This happens because the luminosity and velocity dispersion functions obtained from SDSS are different from those in the SMBH sample which define the relations (\citealt{Bernardi-07}; \citealt{Tundo-07}). \citet{Bernardi-07} argue that the bias lies in the SMBH sample. Assuming that it is due to a selection effect, their models suggest that the \msig\ relation is more fundamental. \citet{Lauer-07} who examine a sample of brightest cluster galaxies, however, find that the local SMBH mass function for \mbh\,$>3\times10^9$\msun\ inferred from the \mlum\ relation is in better agreement with the volume density of the most luminous quasars. Furthermore, from the quasar luminosity function, \citet{Shields-06} estimate the density of relic SMBH with masses $>5\times 10^9$\msun\ to be $~10^{2.3}$Gpc$^{-3}$. This should translate to the same density for galaxies with \sig\,$\gtrsim500$ km s$^{-1}$ based on the local \msig\ relation. No objects with such high \sig\ have been found in the local universe. \citet{Bernardi-06} find at most two or three candidates in SDSS with \sig\,$>500$ km s$^{-1}$ in a volume of $\sim0.5$ Gpc$^{-3}$, but these might be a result of superposition effects. 

Several other authors (\citealt{Netzer-03}; \citealt{Wyithe-03}; \citealt{Shields-04}) have also noted that the \sig\ implied (via the \msig\ relation) by the largest SMBH masses inferred from quasars exceed the largest \sig\ found in local galaxies. If we trust that those SMBH masses are correct, then the \msig\ relation must be different at the upper end. \citet{Wyithe-06} argues that the \msig\ relation is curved upwards at the high-\sig\ end rather than linear in log-log space. If this is true, then the abundance of local SMBHs would be closer to the quasar prediction.

Since the shape of the \msig\ relation at the upper end critically determines the space density of the most massive SMBHs, it is important to characterise its slope and intrinsic scatter. Currently, this high-\sig\ regime is scarcely sampled and there are uncertainties as to how the relation should behave. \citet{Beifiori-09} derive upper limits of 105 SMBH masses based on HST spectroscopy of ionised gas and find that the \msig\ relation flattens at the high-\sig\ end, opposite to what was suggested by \citet{Wyithe-06}. To resolve these uncertainties, an increased sample of direct SMBH detections in high-$\sigma$ galaxies is required. If the \msig\ relation indeed breaks down in the high-\sig\ regime, then a sufficiently large number of SMBH mass measurements at \sig$\sim300-400$ km s$^{-1}$ will likely be able to detect this. For this reason, we undertake an observational campaign to measure SMBH masses of high-\sig\ galaxies using the near-infrared integral-field spectrograph SINFONI at the Very Large Telescope (VLT).

NGC 1332 is a nearby massive S0 galaxy which resides in the Eridanus Cloud. Its orientation is close to edge-on and the galaxy appears to be a normal lenticular. Adopting the bulge velocity dispersion from HyperLeda\footnote{http://leda.univ-lyon1.fr/} ($\sim320$ km s$^{-1}$), we expect that the sphere of influence has a diameter of $\sim0.76$ arcsec, which is resolved by our observations. The $K$-band magnitude of NGC 1332 from the 2MASS Large Galaxy Atlas is 7.052, which gives a K-band luminosity of $1.56\times10^{11}$\lsun\ after a correction for Milky Way extinction of 0.012 magnitudes (from the NASA/IPAC Extragalactic Database -- NED). Our photometric bulge--disc decomposition (Section \ref{bulgediscdecomposition}) implies a bulge-to-total luminosity ratio of 0.44. Given this bulge luminosity, we would expect to find a \mbh\ of $1.37\times10^8$\msun\ from the \mbh-$L_K$ relation of \citet{Marconi-03}. On the other hand, the \msig\ relation of \citet{Tremaine-02} or \citet{Gueltekin-09} constrains the \mbh\ to $\sim9.0\times10^8$\msun\ or $\sim9.7\times10^8$\msun\, respectively. The SMBH masses given by these two relations clearly differ by almost an order of magnitude which makes NGC 1332 a particularly attractive case. As a side note, the software and sources that we used to derive the bulge luminosity were different from those used by \citet{Marconi-03}, which could introduce systematic errors. It is, however, unlikely that the situation for NGC 1332 would be significantly affected.

To date, the only published \mbh\ measurement for NGC 1332 is provided by a recent X-ray study of \citet{Humphrey-09} -- hereafter H09. They make use of the {\it Chandra X-ray Observatory} data and rely on the assumption of hydrostatic equilibrium in the analysis, which results in a SMBH mass of $5.2^{+4.1}_{-2.8}\times 10^8$\msun (with the \msig\ relation as a Bayesian prior). This estimate lies in between the prediction of both relations, although considering the error bars, the \msig\ relation is slightly favoured.  It is, however, not clear if their black hole mass would lean more towards the \mlum\ relation if they were to use the \mlum\ instead of the \msig\ relation for their Bayesian prior. 

In this paper, we measure the SMBH mass in NGC 1332 using a stellar dynamics approach which is not biased by any of the two relations. Throughout, we adopt a distance of 22.3 Mpc from \citet{Tonry-01}, corrected for the Cepheid zero point by applying a distance modulus shift of  -0.06 magnitudes \citep{Mei-05}. At this distance, 1 arcsec corresponds to 0.11 kpc. 

This paper proceeds as follows. We describe the data and data reduction in Section \ref{data}. Details on the derivation of the kinematics follow in Section \ref{kinematics}. The photometry, bulge--disc decomposition and deprojection to model the luminosity profile are described in Section \ref{structureandluminositymodelling}. In Sections \ref{dynamicalmodelling} and \ref{results}, we present the dynamical modelling and the results. Lastly, we summarise and discuss our results in Section \ref{discussions}.

\section[]{Data}
\label{data}

\subsection{SINFONI observations and data reduction}
\label{sinfoniobservations}
The integral-field data presented here were obtained on November 25, 2008, as part of the guaranteed time observations with SINFONI on the UT4 of the VLT. SINFONI (Spectrograph for INtegral Field Observations in the Near Infrared) is a Spectrometer for Infrared Faint Field Imaging (SPIFFI) \citep{Eisenhauer-03a} combined with the Multi-Application Curvature Adaptive Optics (MACAO) module \citep{Bonnet-04}. NGC\, 1332 was observed in the $K$-band (1.95-2.45 $\mu$m) in two different spatial resolutions, i.e. $0.05\times0.1$~arcsec$^2$~spaxel$^{-1}$ (hereafter "100mas") resulting in a $3\times3$~arcsec$^2$ field-of-view (FOV) and $0.125\times0.25$~arcsec$^2$~spaxel$^{-1}$ with $8\times8$~arcsec$^2$ FOV (hereafter "250mas"). For the former, adaptive optics (AO) correction was applied using the nucleus of the galaxy as the natural guide star. The observations followed a sequence of 10-minute exposures of object-sky-object-object-sky-object (O-S-O-O-S-O). Each exposure was dithered by a few spaxels to allow for bad pixel correction and cosmic ray removal. A total of 40 minutes on-source exposure time was obtained for the 250mas scale and 80 minutes for the 100mas scale. To estimate the AO performance and the point spread function (PSF) due to atmospheric turbulence, we regularly observed a PSF star after the science exposure sequence.

\begin{figure}
  \centering
  \includegraphics[scale=0.3]{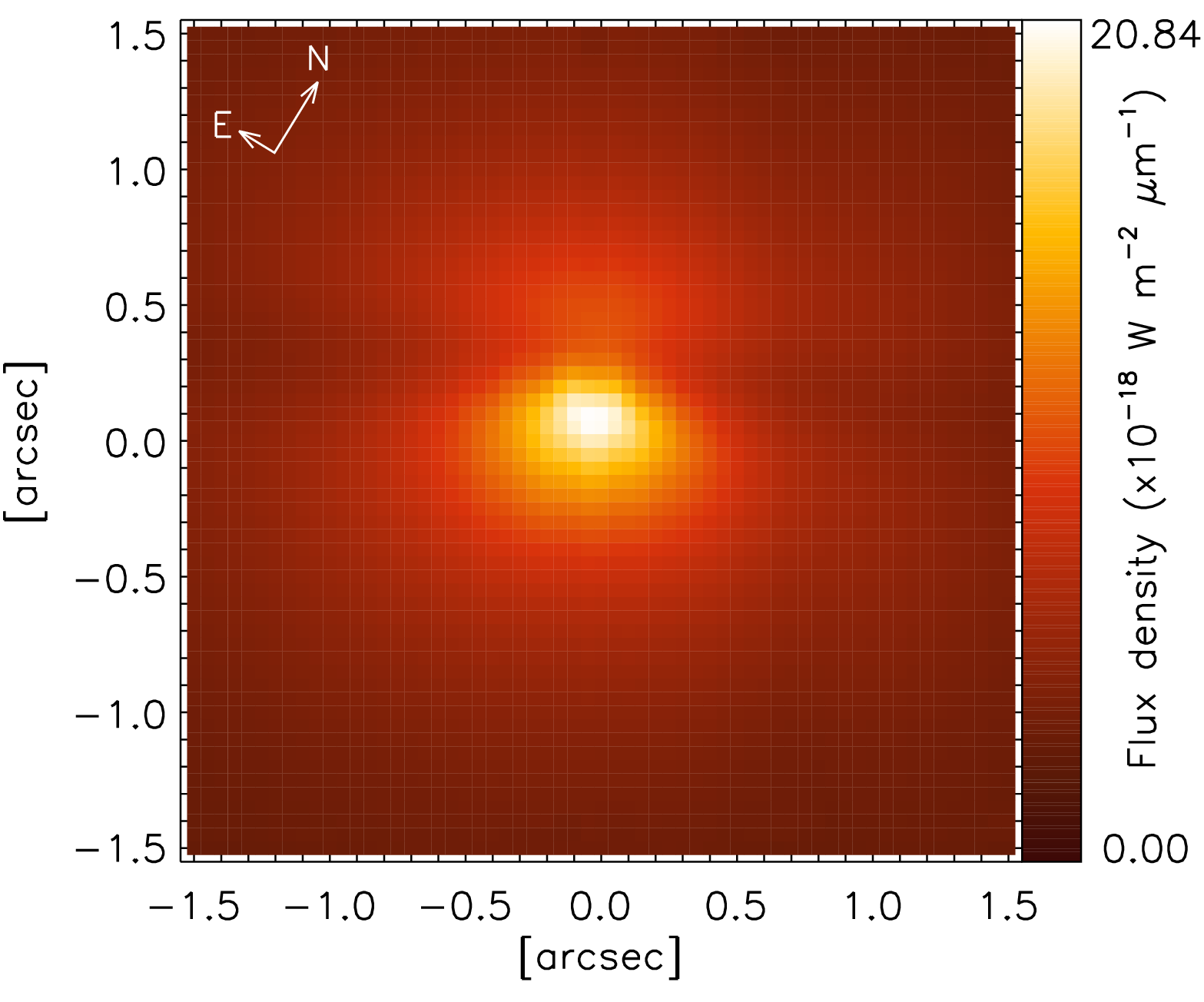} \includegraphics[scale=0.3]{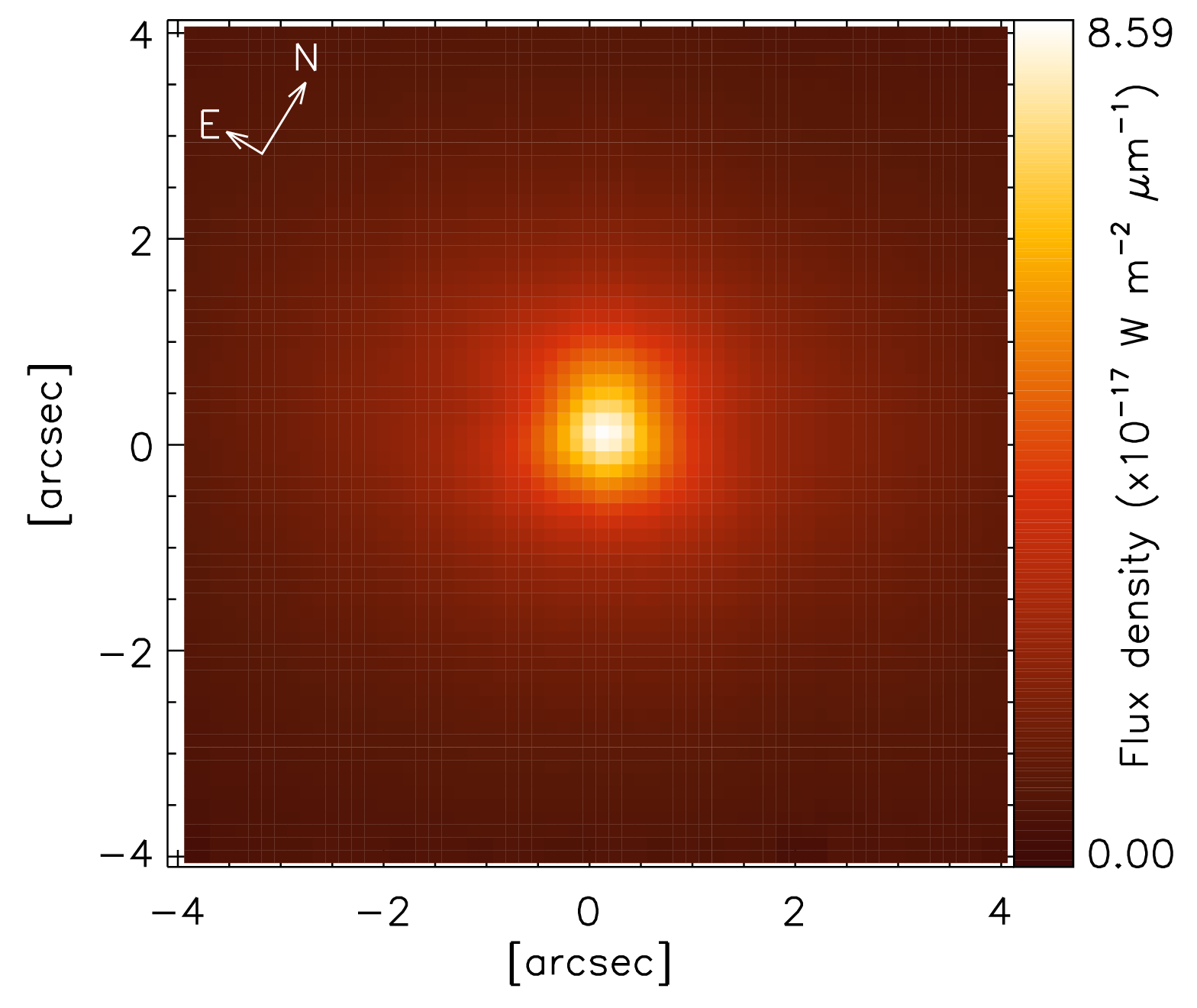}
  \caption{SINFONI images of NGC 1332 in two resolution scales: 100mas (upper panel) and 250mas (lower panel).}
  \label{sinfoimages}
\end{figure}

The reduction of SINFONI data was performed using custom reduction packages for SINFONI, i.e. {\sc ESOREX} \citep{Modigliani-07} and {\sc SPRED} (\citealt{Schreiber-04}; \citealt{Abuter-06}). Except for the 250mas science data, all other frames including the PSF and telluric standard stars were reduced using {\sc ESOREX} which produced cleaner spectra. Both software packages included all common and necessary steps to reduce three-dimensional data and to reconstruct a datacube. The closest sky frame was first subtracted from the science frame. The resulting frame was then flat-fielded, corrected for bad pixels and detector distortion and wavelength calibrated before the datacube was reconstructed. For telluric correction, we used three early-type stars with the spectral classes B3V and B5V, i.e. Hip014898, Hip023060 and Hip018926. As the end process, the individual science datacubes were averaged into one final three-dimensional datacube per plate scale. For a more detailed description of the data reduction, we refer the reader to \citet{Nowak-08}. Fig. \ref{sinfoimages} shows SINFONI images of the two platescales which resulted from collapsing the datacubes along the wavelength direction.

\begin{figure}
  \centering
  \includegraphics[scale=0.48]{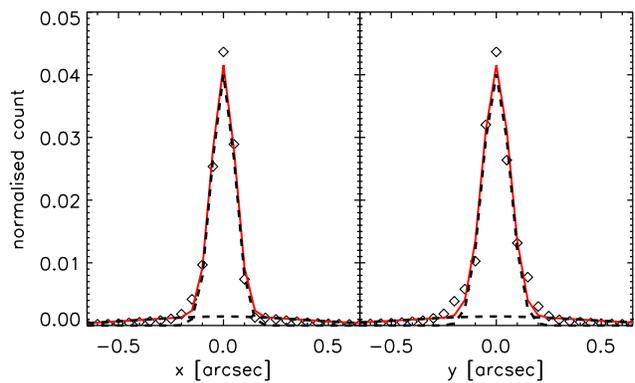}
  \caption{A double non-circular Gaussian fit to the SINFONI 100mas PSF. The fit along the x and the y-axis of the detector are shown in the left and the right panel respectively. The black dashed lines are the individual Gaussians and the red solid lines are the sum of the Gaussian components; diamonds are the actual PSF.}
  \label{sinfopsf}
\end{figure}

After the reduction, the individual PSF star images for the 100mas scale were averaged, normalised and then fitted with the commonly-adopted double Gaussian function. To account for the asymmetry of the PSF, both Gaussian components, i.e. a broad and a narrow one, were set to be non-circular. The fit gave the full width at half-maximum in the x and y direction (${\rm FWHM_x}$, ${\rm FWHM_y}$) of (1.03 arcsec, 0.94 arcsec) for the former and (0.15 arcsec, 0.13 arcsec) for the latter. The narrow component contributed to 36\% of the total flux and resolved the expected sphere of influence of the SMBH. We used this SINFONI PSF parameterisation for the surface brightness deprojection (see Section \ref{deprojection}). Fig.~\ref{sinfopsf} presents the PSF image with the fit along the two spatial axes. We note that this PSF image does not strictly represent the true PSF since the acquisition was not done simultaneously with the galaxy observation. The fit is also admittedly not perfect, but for our purpose here the deviation does not lead to a significant error. We verify this and discuss the effect of the PSF uncertainties further in Appendix \ref{appendix}.

\subsection{Imaging data}
\label{imagingdata}
\begin{figure*}
  \centering
  \includegraphics[scale=0.75]{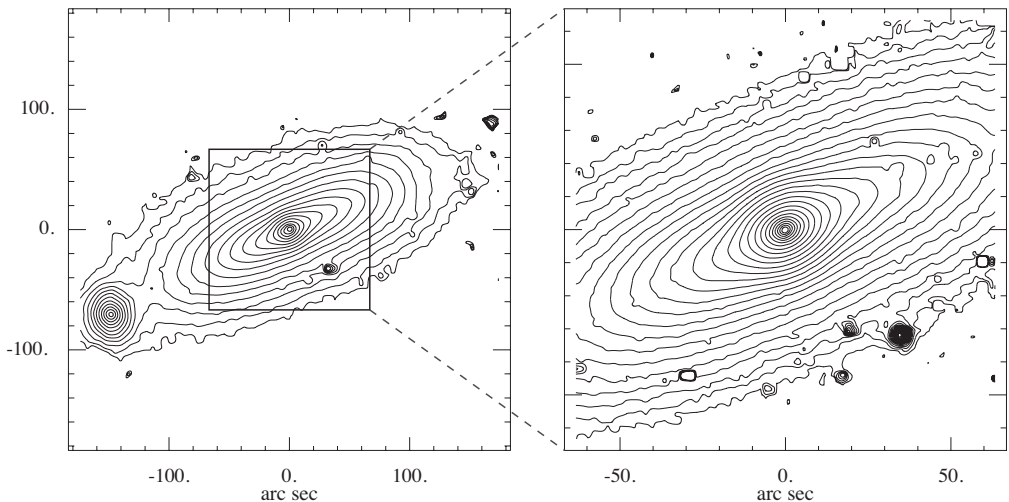}
  \caption{$R$-band isophotes of NGC 1332, from archival NTT-EMMI images. The left-hand panel shows the whole galaxy, plus its neighbour NGC 1331 (near the left-hand edge); the image has been smoothed with a 5-pixel-wide median filter.  Contour levels run from 24.5 to 15.0, in steps of 0.5 mag/arcsec$^{2}$. The right-hand panel shows the inner regions of the galaxy, including the bulge, using an unsmoothed image; contour levels run from 23.1 to 15.0, in steps of 0.3 mag/arcsec$^{2}$. North is up and east is to the left.}
  \label{contour}
\end{figure*}

We made use of three types of imaging data for NGC 1332. For the large-scale analysis, we searched the major telescope archives and found several $R$-band images obtained with the red channel of ESO Multi-Mode Instrument (EMMI) on the 3.5m New Technology Telescope (NTT) at La Silla.  These images were originally taken on 2005 December 3, as part of a spectroscopic program (Program ID 076.B-0182(A), PI Arag\'{o}n-Salamanca). We selected the three best 10s exposures (the fourth exposure had strong background variations) and reduced them with standard IRAF tasks (first reducing the individual amplifier sections, then scaling and joining them into single-chip images).  Since the imager was actually a two-CCD mosaic, the result was three pairs of single-chip images; we combined these into a single mosaic using the SWarp package \citep{Bertin-02}. The final image had a seeing of 0.80 arcsec FWHM (mean of Moffat profiles fitted to nine bright, unsaturated stars) and a plate scale of 0.332 arcsec pixel$^{-1}$, and is shown in Fig.~\ref{contour}.

From the HST archive we retrieved two 160s Wide-Field Planetary Camera 2 (WFPC2) images in the F814W filter (Proposal ID 5999, PI Phillips), which were also used in \citet{Kundu-01}. Since the images were well aligned, we combined them directly using the STSDAS crrej task in IRAF. Finally, we generated $K$-band images from our 250mas and 100mas SINFONI datacubes by collapsing (averaging) the cubes along the spectral direction.

We first calibrated the combined NTT-EMMI mosaic image, which was large enough to ensure accurate sky subtraction, using Cousins $R$ photometry from the literature \citep{Prugniel-98}.  The WFPC2 image was then calibrated by matching surface brightness profiles: fitting ellipses of fixed orientation (position angle and ellipticity) to both images, then simultaneously fitting for the best combination of sky background (in the WFPC2 image) and scaling so that it matched the NTT-EMMI profile outside the central 2 arcsec (where the worse seeing of the NTT-EMMI image affected the profile).  Finally, we repeated the process for profiles from the two SINFONI images by matching them against the (scaled, sky-subtracted) WFPC2 profile.

\section[]{Kinematics}
\label{kinematics}
\subsection[]{SINFONI kinematics}
\label{sinfonikinematics}
We extracted the line-of-sight velocity distributions (LOSVDs) non-parametrically using a Maximum Penalised Likelihood (MPL) method \citep{Gebhardt-00}. The galaxy spectra were deconvolved using the weighted linear combination of a set of stellar templates consisting of K and M stars. These stars were previously observed using SINFONI with the same instrumental setups as in the galaxy observations. We briefly describe here the kinematics analysis that we have performed. It is largely similar to that in \citet{Nowak-07} and \citet{Nowak-08}, so we refer the reader to those papers for details on the kinematics derivation. 

Reliable kinematics from MPL can be obtained when the S/N is sufficiently high. Therefore, to optimise and homogenise the S/N, we binned the pixels into angular and radial bins as in \citet{Gebhardt-03} by luminosity-weighted averaging of the spectra. The galaxy is divided into four quadrants bordered by the major and minor axes. Each quadrant is divided further into five angular bins. The centres of those bins are at the angles of  5.8\degree, 17.6\degree, 30.2\degree, 45.0\degree and 71.6\degree. For the 100mas data, there are seven or eight radial bins while for the 250mas data, 12-13 of those bins were needed to cover the FOV. We then performed the MPL method on the binned spectra as follows. 

We first normalised the galaxy and stellar template spectra by dividing out the continua. The combined stellar template was convolved with a binned initial LOSVD. The LOSVD and the weights of the templates were iteratively changed until the convolved combined spectrum matched the galaxy spectrum. This fit was done by minimising the penalised $\chi^2$: $\chi_p^2 = \chi^2+\alpha P$. A certain level of smoothing was applied to the LOSVD via the second term where $P$, the penalty function, is the integral of squared second derivative of the LOSVD. The smoothing parameter $\alpha$ determines the level of regularisation and its value depends on the velocity dispersion of the galaxy and the S/N of the data. We estimated the appropriate smoothing for our data from the kinematic analysis of a large dataset of model galaxy spectra. These models were created by broadening the template spectrum with a velocity dispersion of 400 km s$^{-1}$. Our data have a high S/N which reaches $~70$ in the central pixel. After the binning, the S/N increased to $\sim90$ (100mas) and $\sim83$ (250mas) on average. With those S/N values, the appropriate values for $\alpha$ found in the above simulations are on average $\sim5$ and $\sim6$, respectively. 

To derive the LOSVDs, we specifically fitted the first two CO bandheads  CO(2-0) and CO(3-1) in our spectra. To minimise the error due to template mismatch, we measured the equivalent width of the first CO bandhead as in \citet{Silge-03} and selected only stars with similar equivalent widths for the templates. Across the FOV of SINFONI 250mas, the measured values range from $\sim11$ to $\sim15$\AA.

We calculated uncertainties for each LOSVD from 100 Monte Carlo realisations of the galaxy spectra. These spectra were obtained by convolving the measured LOSVD with the stellar templates. Each spectrum differs from the others in the amount of Gaussian noise added to the spectrum at each wavelength position. From every spectrum, an LOSVD was extracted and used to estimate the errors. For illustration purposes, we parametrised the LOSVDs in terms of Gauss-Hermite moments (\citealt{vanderMarel-93}; \citealt{Gerhard-93}), i.e. velocity $v$, velocity dispersion $\sigma$ and two higher order terms which measure the asymmetric and symmetric departure from a pure gaussian velocity profile $h_3$ and $h_4$. The typical errors that were derived for $v$ and $\sigma$ are 8.45 km s$^{-1}$ and 8.87 km s$^{-1}$ (100mas), respectively; for 250mas, the errors are 7.92 km s$^{-1}$ and 9.21 km s$^{-1}$. For $h_3$ and $h_4$, the errors are typically 0.02 for both scales.

\begin{figure*}
  \centering
  \includegraphics[scale=0.212]{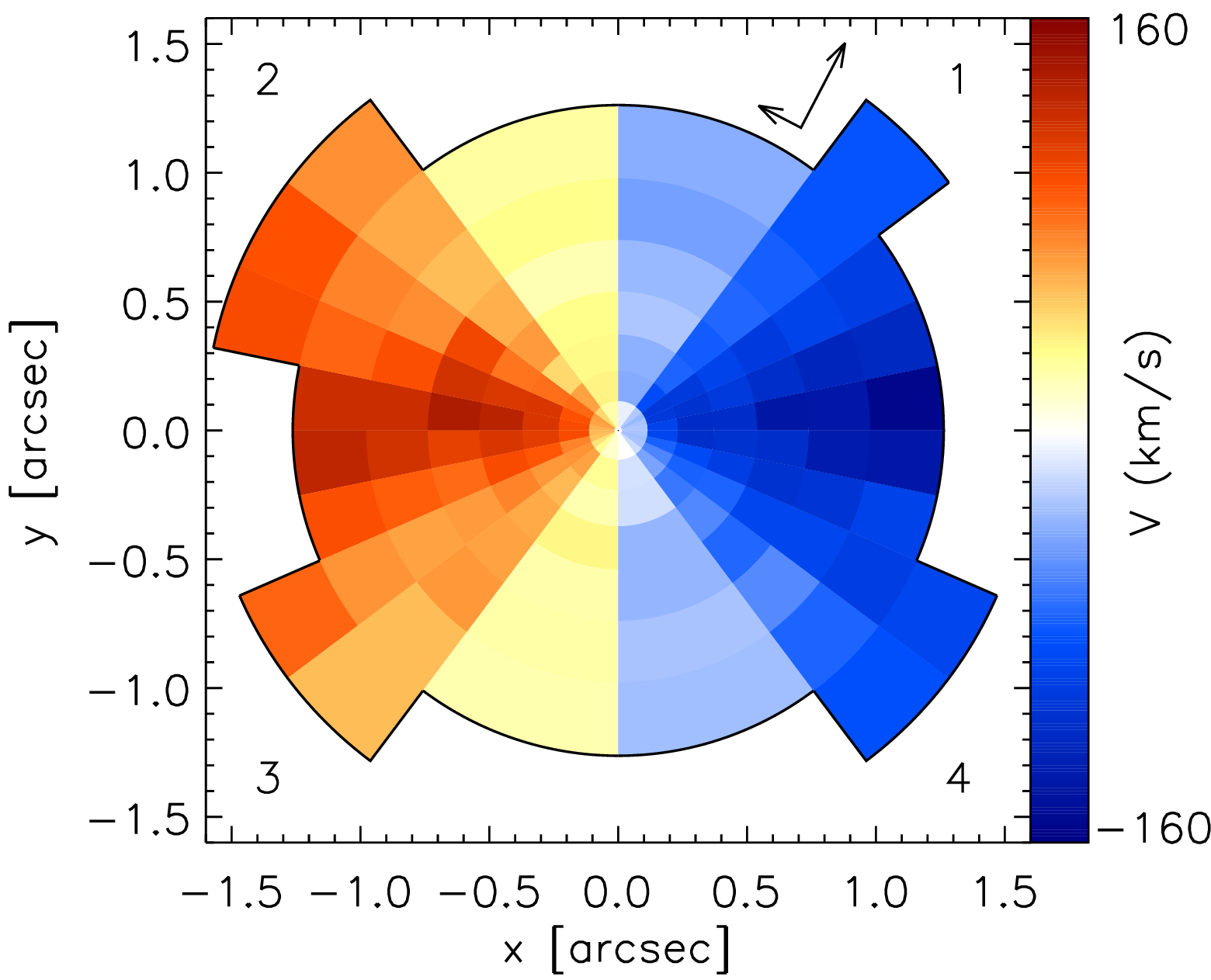} \includegraphics[scale=0.212]{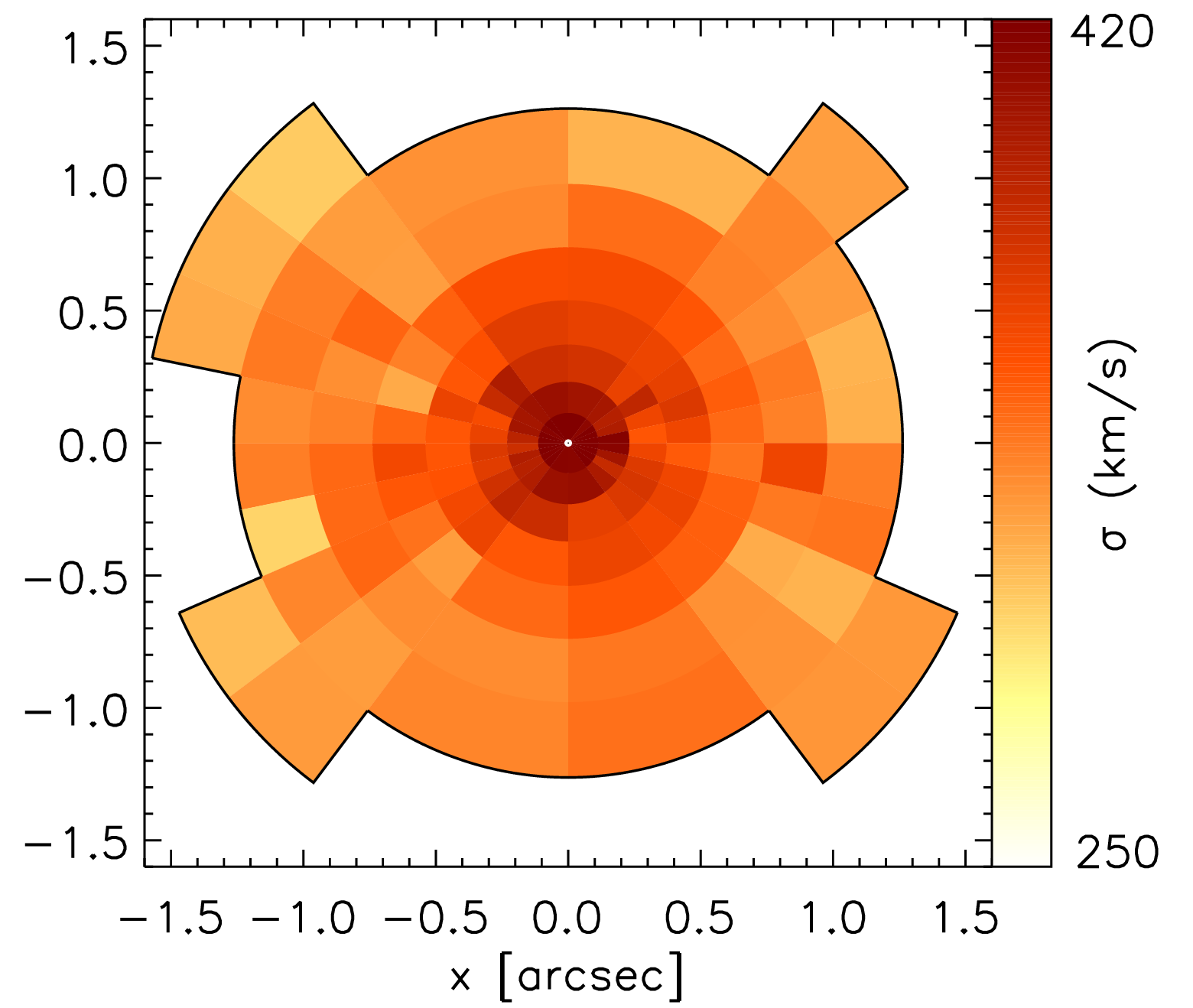} \includegraphics[scale=0.212]{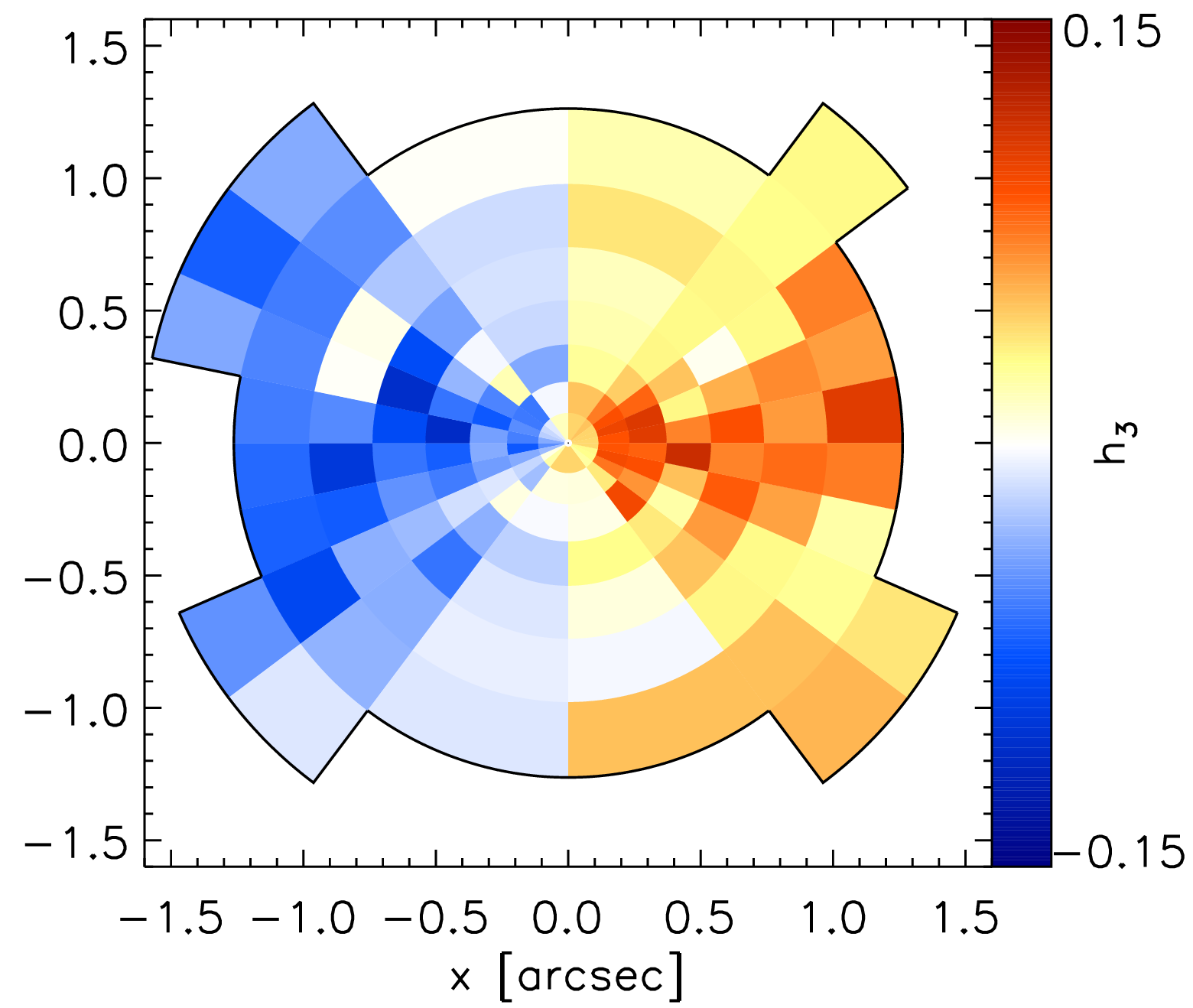} \includegraphics[scale=0.212]{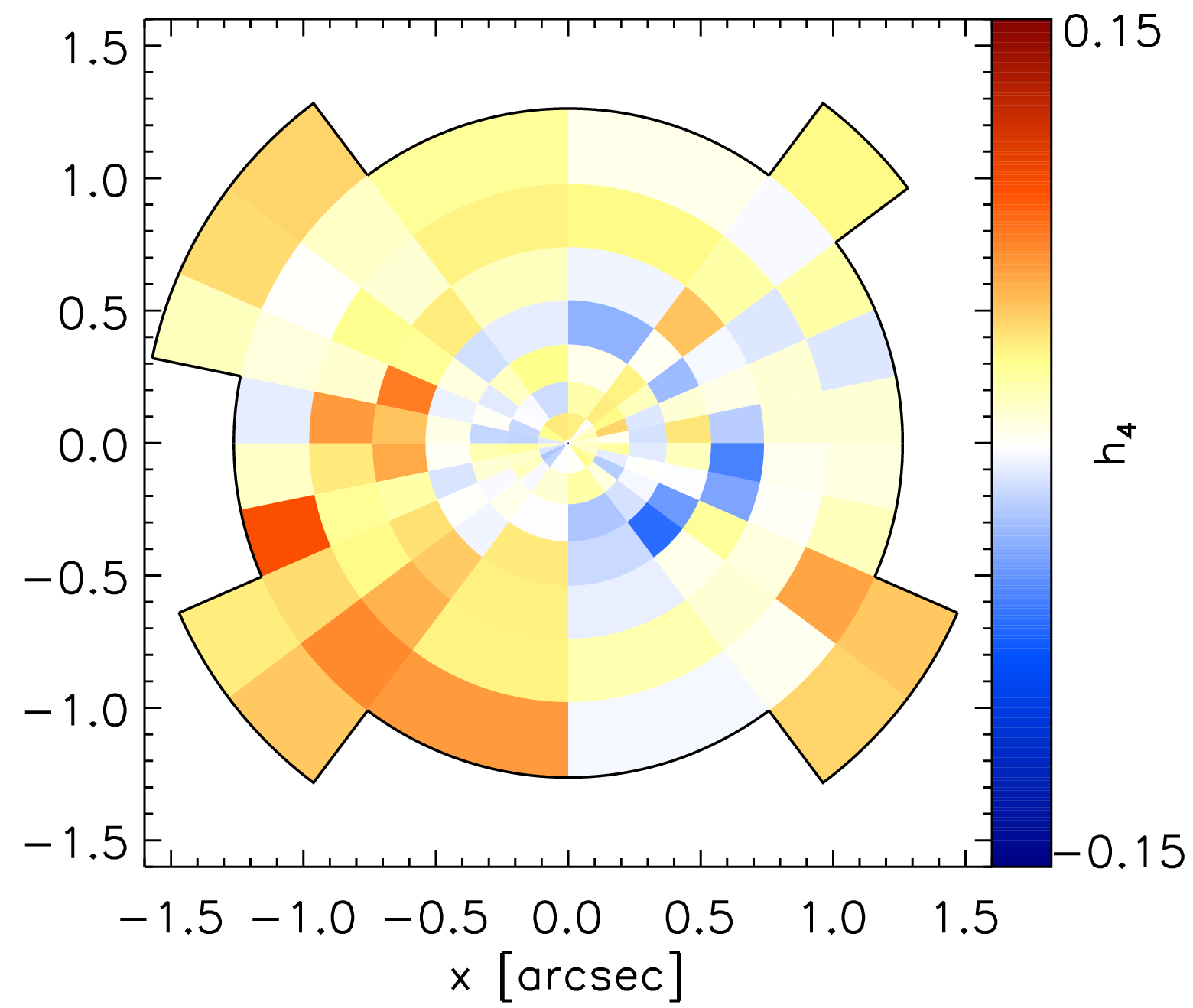}
  \includegraphics[scale=0.212]{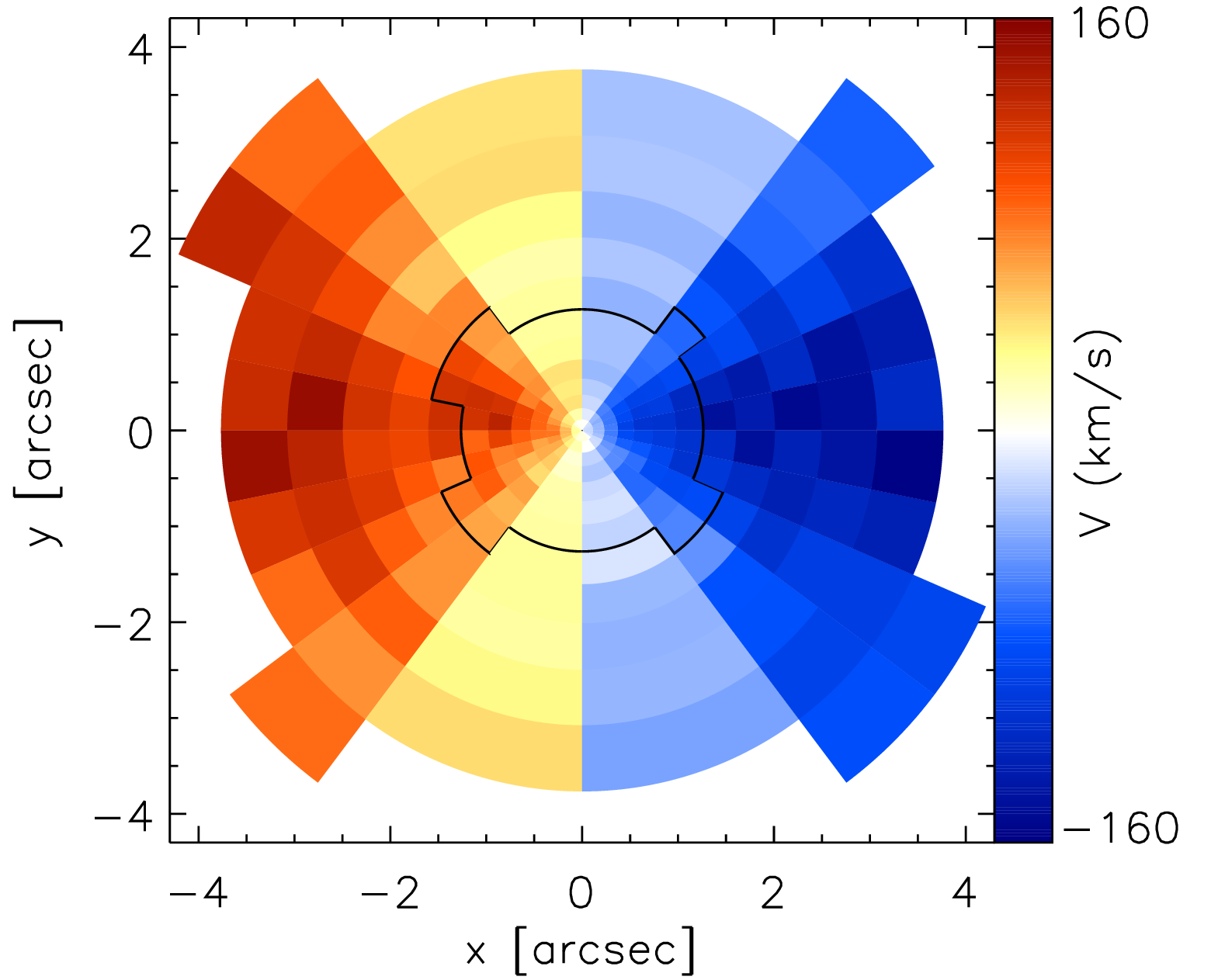} \includegraphics[scale=0.212]{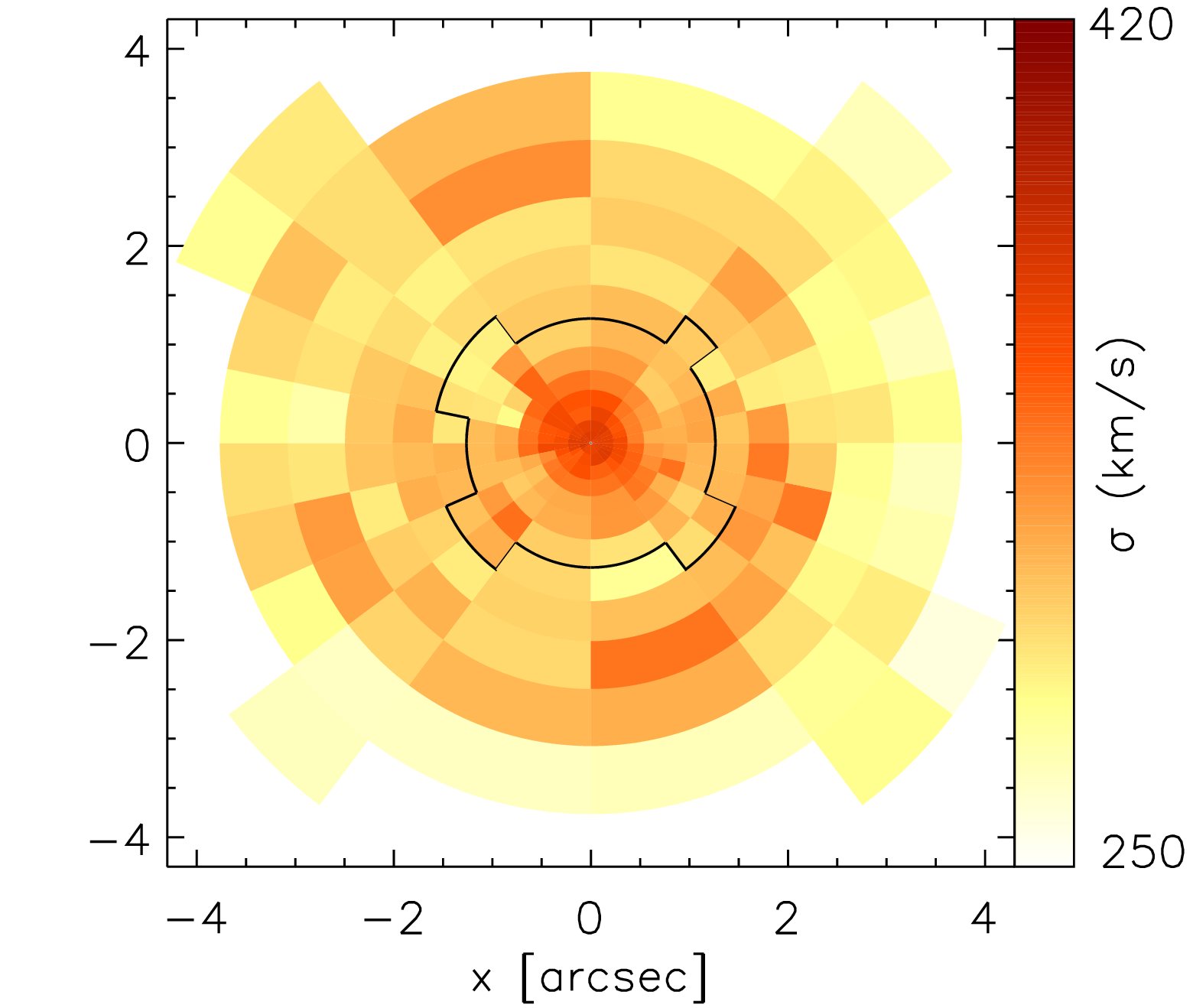} \includegraphics[scale=0.212]{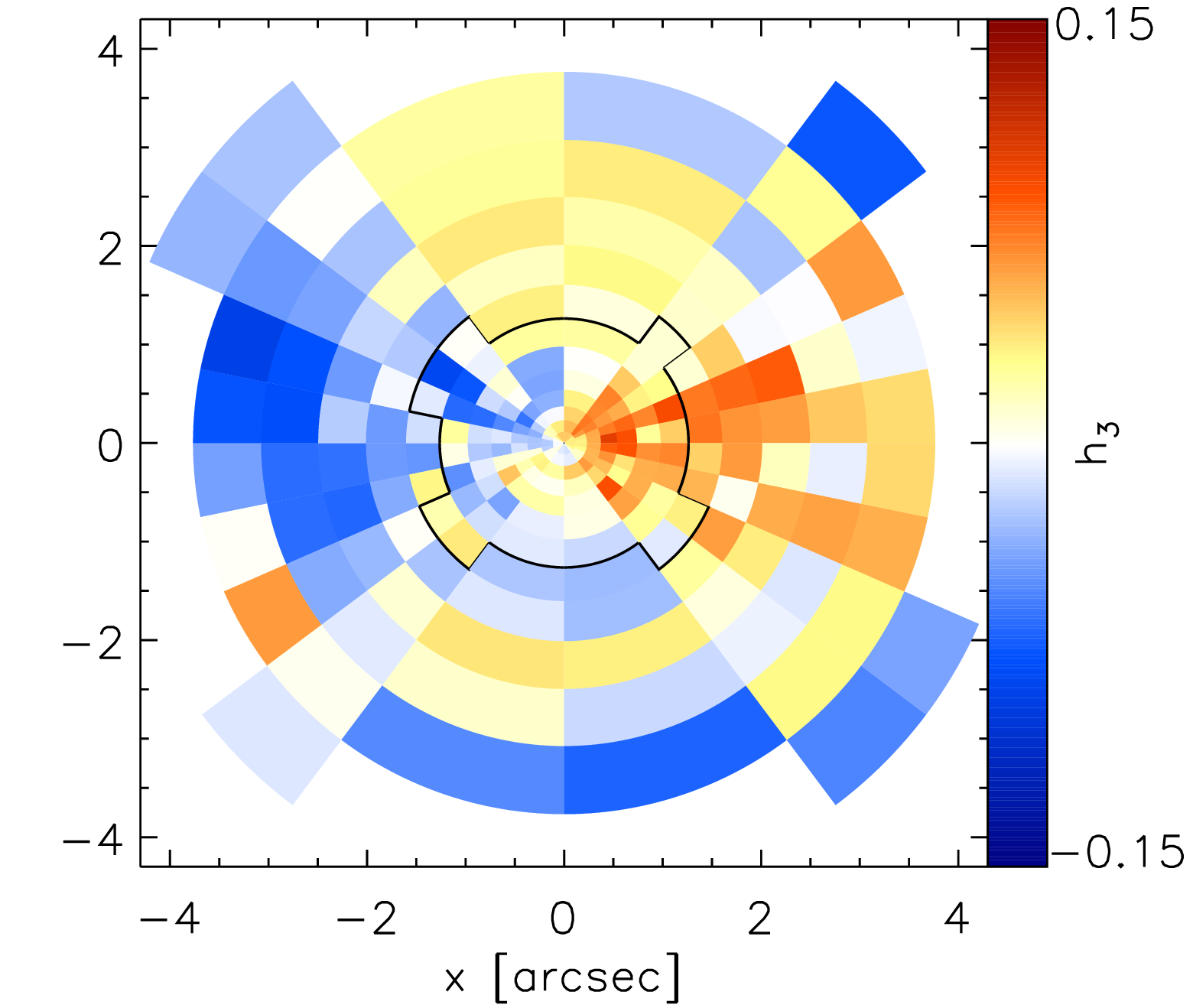} \includegraphics[scale=0.212]{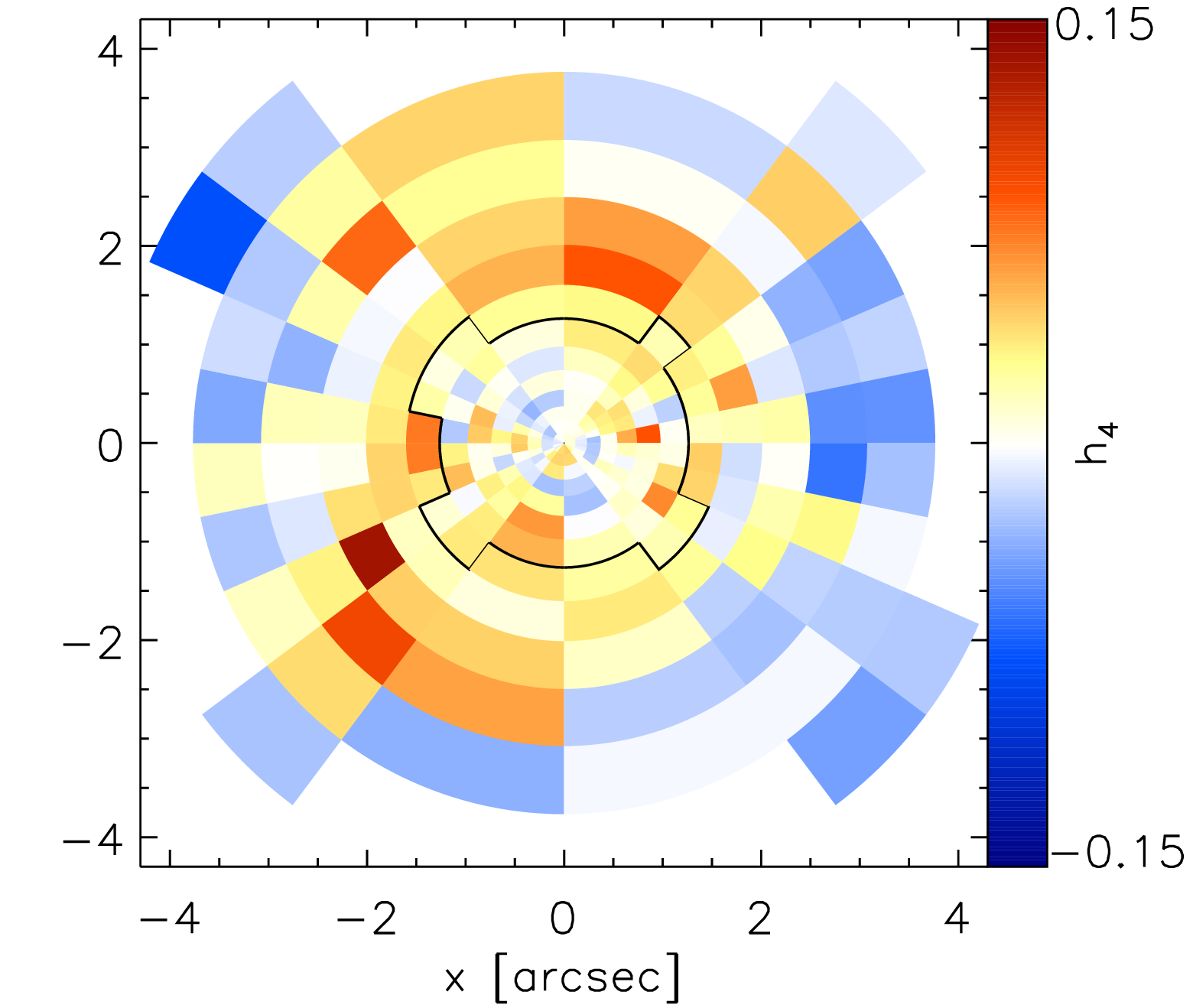}
  \caption{Kinematics of NGC 1332 derived from SINFONI data, illustrated in terms of Gauss-Hermite moments (v, $\sigma$, $h_3$ and $h_4$) as written on the side of each map. The first row displays 100mas data while the second row shows 250mas scale kinematics in which the spatial scope of the 100mas data is outlined. The map shows the division into radial and angular bins as described in the text. The major and minor axes of the galaxy are aligned with the abscissa and the ordinate of the coordinate system respectively. In the first map, the numbers (1-4) correspond to the quadrant numbers and the arrows indicate the orientation to the north (long) and the east (short).}
  \label{kinematicsmap}
\end{figure*}

We present kinematic maps of NGC 1332 for both scales in Fig.~\ref{kinematicsmap}. A significant rotation is shown by the well-ordered pattern in the velocity map and the anti-correlating $h_3$. The velocity dispersion is peaked at around 400 km s$^{-1}$ and there is a rather steep decline towards the outskirts. In spite of the presence of dust in the nucleus, the kinematic centre seems to coincide with the photometric centre.    

\subsection[]{Long-slit kinematics}
\label{longslitkinematics}
Our SINFONI data provide the required resolution to allow for an accurate measurement of the SMBH mass at the expense of the FOV size. To obtain constraints on the orbital structure at larger radii, we utilised long-slit data which were reported in \citet{Kuijken-96}-- hereafter K96. They observed NGC 1332 using the Red Channel Spectrograph at the Multiple Mirror Telescope. The optical data were taken along the major axis of NGC 1332 through a 1.25 arcsec$\times$3 arcmin slit with a spectral resolution of 2.6\AA\ (instrumental \sig=63 km s$^{-1}$). The major axis position angle (PA) given in K96 is 148\degree\ following RC3. This is about 30\degree\ higher than the PA we measured from the WFPC2 image, i.e. 117\degree. The latter PA is consistent with our SINFONI data, and also with the PA given by HyperLeda and ESO/Uppsala Survey. By simply looking at images of NGC 1332, a PA of 148\degree\ can be easily rejected. We have confirmed that the slit was indeed placed along the long axis of the galaxy and therefore the quoted PA in K96 is a typo (K. Kuijken, private communication).  

The velocity profile of the slit data was derived using the Fourier Correlation Quotient (FCQ) method \citep{Bender-90}, parametrised into Gauss-Hermite parameters $v$, \sig, $h_3$, $h_4$. The LOSVDs were then reconstructed from those moments as it is the full LOSVD that is fitted in the modelling. The data provide kinematics up to a radius of $\sim80$ arcsec on both sides of the major axis.

The SINFONI and the long-slit datasets overlap within 3.5 arcsec and they are broadly consistent with each other at radii $\gtrsim 2$ arcsec. Inside that radius, $v$ and $\sigma$ derived from the long-slit data are systematically lower. For instance, at $\sim 0.2$ arcsec, the slit data give a \sig\ of $\sim335$ km s$^{-1}$ while the 250mas and the 100mas SINFONI data give $\sim365$ km s$^{-1}$ and $\sim385$ km s$^{-1}$ respectively. Moreover, the SINFONI \sig\ also shows a steeper gradient. We suspect that these differences are due to the seeing during the slit observation and the better resolution of the SINFONI data. Since the PSF of the slit observation is not known, it is difficult to investigate the discrepancy. However, we find that within the discrepant region SINFONI kinematics dominates and renders the slit kinematics unimportant, as briefly explained in the following. We modelled NGC 1332, as described in Section \ref{dynamicalmodelling}, using both datasets simultaneously. We ran several models using identical set-ups, varying only the slit PSF to several reasonable values (1.5 arcsec and 2 arcsec). We also ran models where we used only slit data outside the SINFONI FOV ($>3.5$ arcsec). We are able to show that all those variations do not alter the modelling results. With this finding, whenever we include slit kinematics in the dynamical modelling, we consider only slit datapoints outside the SINFONI FOV to ease the computational load and time.

\section[]{Structural and Luminosity Modelling}
\label{structureandluminositymodelling}
\subsection{Bulge--disc decomposition}
\label{bulgediscdecomposition}
Since NGC 1332 is an S0 galaxy with a fairly well-defined bulge and disc, we investigated bulge-disc decompositions, both for modelling purposes (i.e., in case the bulge and disc stellar populations might have different mass-to-light ratios) and so that we could estimate its bulge luminosity in order to see where NGC 1332's SMBH fell in the SMBH-bulge relations. We tried two approaches, which agreed extremely well.  The first involved fitting free ellipses (i.e., with position angle and ellipticity allowed to vary) to the isophotes of the NTT-EMMI and SINFONI images (carefully masking out the dust lane in the SINFONI images) and then combining these into a single 1-D surface-brightness profile (see Section \ref{photometricmodels}).  The best fit to this profile was a combination of an exponential (representing the outer disc, with central surface brightness $\mu_{0} = 18.78$ and scale length $h = 33.6$ arcsec), a S\'ersic function (representing the bulge, with $n = 2.36$, surface brightness $\mu_{e} = 18.36$, and effective radius $r_e = 9.15$ arcsec), and a small, narrow Gaussian representing a possible central star cluster.

We also tried a 2D bulge--disc decomposition, using version 2.2 of the BUDDA image-fitting code \citep{deSouza-04,Gadotti-08} and the NTT-EMMI image by itself; we masked the dust-affected part of the nucleus and turned on seeing correction.  The best fit was with an exponential disc having an ellipticity of 0.73 ($\mu_0 = 18.66$, $h = 32.96$ arcsec), a S\'ersic bulge with ellipticity = 0.27 ($n = 2.34$, $\mu_e = 18.18$, $r_{e} = 8.39$ arcsec), and a small point source (representing, e.g., a stellar nucleus) contributing to 1.2\% of the total light.  The agreement with the 1-D decomposition is excellent. We note that the 2D decomposition corresponds to a bulge-to-total light ratio of 0.43 (0.44 if we include the small point source), so NGC 1332 is, despite its high central velocity dispersion, still a (marginally) disc-dominated galaxy.

\subsection{Photometric models}
\label{photometricmodels}
We constructed two main photometric models. The first was a single-component model, where the galaxy was represented by a single surface-brightness profile, with variable ellipticity and higher-order moments $a_{4}$, $a_{6}$, etc. \citep{Bender-87}.  To construct this profile we fit ellipses to isophotes using three images: the NTT-EMMI $R$-band image and the two SINFONI images (i.e., 100mas and 250mas datacubes collapsed along the wavelength axis). For both SINFONI images, we masked out the regions affected by the nuclear dust lane before fitting ellipses. The NTT-EMMI profile was used for $r > 4.5$ arcsec; at smaller radii, where better resolution and lower dust extinction were needed, we used the SINFONI profiles (the 250mas profile was used from 4.5 arcsec in to 1.5 arcsec, with the profile at $r < 1.5$ arcsec coming from the 100mas image). We see no noticeable gradient in $V-I$ colour from HST images in the innermost 13 arcsec along the south semi-minor axis, which is less affected by dust. Since dust regions were masked out during profile extraction, $V-I$ should be flat, and so we do not expect a $R-K$ color variation, either. The combination of images from the $R$ and $K$ bands can therefore be justified.

The second photometric model involved separating the galaxy into a bulge and a disc, with potentially different stellar mass-to-light ratios.  Since the 2D bulge-disc decomposition worked well, we decided to use the disc component from that fit as the disc component for modeling purposes.  Because we wanted to match the actual surface brightness and isophote shapes in the central regions as closely as possible, using the (fixed-ellipticity) S\'ersic bulge model from the 2D fit would have been too crude an approximation; even including the small Gaussian component still produces residuals and does not reproduce the actual ellipticity profile of the galaxy's inner regions.  Instead, we first subtracted the disc model from our images (specifically, the NTT-EMMI image and the two SINFONI images) to create residual, ``bulge-only'' images and then performed variable-ellipticity fits on these images in order to generate a bulge model for modeling purposes.  The result of this was that the surface brightness and ellipticity of the bulge component tracked the actual brightness and isophote shapes in the very central regions as accurately as possible; moreover, the combined light from this bulge component and the 2D disc model reproduces the original galaxy light distribution.

To construct the final bulge component, the ellipse fits to the NTT-EMMI image were used for $r > 4.5$ arcsec; the 250mas SINFONI image was used for $r = 1$--4.5 arcsec, and the 100mas image was used for $r < 1$ arcsec.  As was the case for the single-component model, we carefully masked out the dust lane in both SINFONI images before running the ellipse-fitting software.  At radii $> 15$ arcsec, the NTT-EMMI profile from the residual image became significantly affected by deviations of the disc from the 2D model which we subtracted from the image. To ensure a relatively smooth luminosity model, we replaced the data at $r > 15$ arcsec with an extrapolation of the best-fitting S\'ersic component from the 2D fits (section \ref{bulgediscdecomposition}), fixing its ellipticity to 0.27. At these radii, the light is dominated by the disc component, so small variations in the bulge component have minimal effect on the modelling.

Finally, we constructed an alternate single-component profile for testing purposes (see Appendix \ref{appendix}); this was identical to the SINFONI+NTT-EMMI profile described above except that ellipse fits to the WFPC2 F814W image were used for $r < 0.5$ arcsec.  Problems with strong dust extinction in the WFPC2 image ultimately led us to reject using this data for the actual modelling.

\begin{figure}
 \centering
 \includegraphics[scale=0.44]{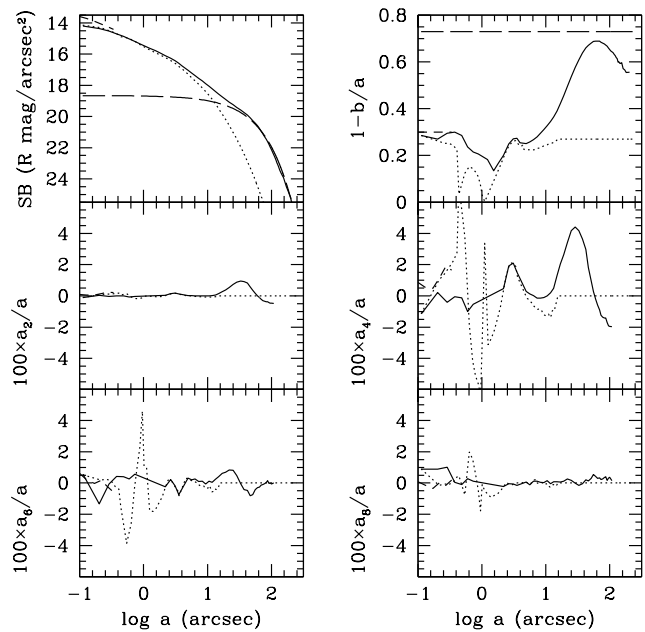}
 \caption{The isophotal shape analysis of NGC 1332. As a function of the logarithm of the semimajor axis distance we show the $R$-band surface brightness (top left), the ellipticity (top right), the $a_2$ (middle left), $a_4$ (middle right), $a_6$ (bottom left) and $a_8$ (bottom right) coefficients of the isophotal Fourier analysis. The solid lines refer to the one-component photometry. The short dashed lines show the photometric parameters as derived from the HST image in the inner 0.5 arcsec. The dotted and long dashed lines show the bulge and disc components of the two-component photometry, respectively.}
 \label{isophote}
\end{figure}

Isophotal shapes of all the models (the one-component model, bulge and disc of the two-component model and the single-component model using WFPC2/HST image) is shown in Fig.~\ref{isophote}.

\subsection{Deprojection}
\label{deprojection}
The dynamical modelling requires knowledge of the three-dimensional luminosity distribution $\nu$ (cf. Section \ref{dynamicalmodelling}). Each photometric profile, i.e., the single-component model, the bulge and the disc model was deprojected separately, resulting in the ($R$-band) 3D luminosity profiles $\nu$, $\nu_b$ and $\nu_d$ respectively. Since the disc flattening is small, the orientation of NGC 1332 must be close to edge-on. We therefore assume an inclination of 90\degree\ throughout the paper unless otherwise stated. The edge-on deprojection gives an intrinsic flattening of 0.27 for the disc.   

The deprojections of the bulge, the disc and the one-component model were done under the assumption of axisymmetry using a modified version of the code of \citet{Magorrian-99}. We briefly describe it as follows. We first constructed an initial density model, defined in equation 10 of \citet{Magorrian-99}. This was done by running through a grid of the model parameters. Each of the models was convolved and projected to be compared with the observed surface brightness. The model with  the smallest $\chi^2$ was selected as the initial model. This model was then refined by applying small changes through a simulated annealing procedure as in \citet{Magorrian-99}. The convolution, projection and comparison steps were subsequently repeated after each change. The iteration was stopped and the final density model was reached when the model matched the observations within a pre-determined accuracy.    

For the disc component, the PSF effect is negligible because the disc is very faint at the innermost radii where the PSF becomes important (the central luminosity density of the disc is at least three orders of magnitude lower than that of the bulge). For the bulge and the one-component model, where the innermost isophotes are based on the SINFONI 100mas image, we had to take the PSF into account during the deprojection. For this purpose, we used the double-Gaussian parameterisation of the SINFONI PSF described in Section \ref{sinfoniobservations} to implement the PSF convolution.

\begin{figure}
  \centering
  \includegraphics[scale=0.48]{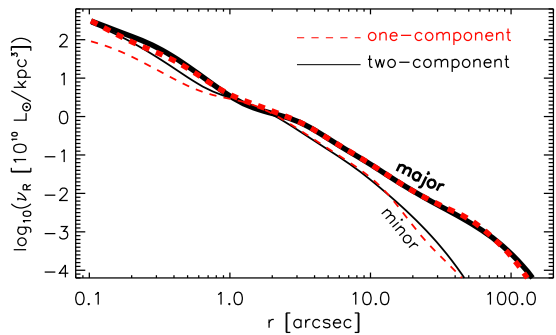}
  \caption{The model of PSF-deconvolved luminosity density of the one-component (red dashed line) and the two-component (black solid line) density models for an inclination of 90\degree. The subscript "R" refers to the $R$-band. Thick and thin lines refer to density profiles along the major and minor axes respectively.}
  \label{lumdens1}
\end{figure}

We show the PSF-deconvolved luminosity density profiles for both density models in Fig.~\ref{lumdens1}. For the two-component models, we plot the total quantity (bulge + disc). The density of the disc exceeds that of the bulge at $r > 12$ arcsec. The difference of the density profiles at $\lesssim$ 1 arcsec reflects a certain level of uncertainties in the photometry and the bulge-disc decomposition, which gets amplified in the deprojection. The luminosity densities differ mainly along the minor axis. It is apparent that the two-component density model is rounder in the centre. We investigate how these uncertainties and differences affect the SMBH mass estimate in Section \ref{results}.

\section[]{Dynamical Modelling}
\label{dynamicalmodelling}
We employed an orbital synthesis method based on \citet{Schwarzschild-79} to model NGC 1332 and to measure the SMBH mass. In particular, we used the three-integral axisymmetric code described in \citet{Gebhardt-00, Gebhardt-03}, \citet{Thomas-04} and \citet{Siopis-09}. The modelling procedure includes the following steps. (1) Calculation of a trial gravitational potential consisting of the contribution from stars and the SMBH. (2) Generation of an orbit library which obeys the given potential. (3) Calculation of orbital weights such that the orbit superposition satisfies the light distribution and the kinematical constraints. (4) Repetition of steps (1)-(3) for different trial potentials obtained by varying the mass-to-light ratio(s) and the SMBH mass. The best-fitting parameters are then found through a $\chi^2$ analysis. 

We used each of the two density models described in Section \ref{deprojection} for the modelling. For the one-component density model, the mass distribution of NGC 1332 followed $\rho=\Upsilon\nu+$\mbh$\delta(r)$, where the stellar mass-to-light ratio \ml\ and \mbh\ were the only free parameters. For the two-component density models, the mass distribution became $\rho=\Upsilon_{b}\nu_b+\Upsilon_{d}\nu_d+$\mbh$\delta(r)$ with the bulge mass-to-light ratio \mlb, the disc mass-to-light ratio \mld\ and \mbh\ as the free parameters. 

Each of our orbit libraries consisted of about $2\times15000$ orbits, i.e. two identical sets of 15000 orbits, opposite in angular momentum directions. We set the maximum radius of the libraries to 100 arcsec and limit the extension of the long-slit data to 30 arcsec. For each of the modelling runs that we conducted (cf. Section \ref{results}), we modelled each quadrant separately, resulting in four different SMBH mass values, one for each quadrant. This highlights the benefit of having integral-field data, i.e. (1)\,to justify the assumption of axisymetry, (2) to have a complete spatial coverage of kinematic data and (3)\,to allow for four independent measurements of the SMBH mass (under the assumption of axial symmetry). In order to implement the PSF convolution in the modelling as accurately as possible, we directly used the SINFONI PSF images from both scales, instead of the double Gaussian parameterisation.

The SINFONI data were mapped into 226 spatial bins (see Fig.~\ref{kinematicsmap}), 120 of which come from the 100mas data and sample the inner part. As kinematic constraints, we used the full LOSVD of each bin, sampled into 25 velocities from -1543 km s$^{-1}$ to 1543 km s$^{-1}$. These provided a total of $226\times25=5650$ kinematic observables to be fitted by the dynamical models (about 1412 observables per quadrant). The LOSVDs from the long slit data (3.5 arcsec$<r<$30 arcsec) were binned in the same way, increasing the total number of observables by 650 ($\sim162$ for each quadrant).

\section[]{Results}
\label{results}
\subsection{The black hole mass and stellar mass-to-light ratio}
As described in the previous sections, we prepared two kinematics datasets and constructed two stellar density models. For the latter, an inclination of 90\degree\ is assumed. In this subsection we discuss the results for the edge-on models. We address the uncertainties due to inclination in another subsection (\ref{inclinationeffects}), where we also show that the edge-on models produce better fits to the data.

We performed five different modelling runs using different combinations of the density models and kinematics to probe the possible systematic uncertainties that could arise from various constraints/datasets. For each density model, there were two runs, one with and one without slit kinematics (100mas and 250mas SINFONI kinematics were always used). We list all runs with the resulting \mbh\ for each quadrant in Table \ref{tablembh} and similarly for the mass-to-light ratios in Table \ref{tablemlb}. The naming of the five runs is described in the caption of the first table. Whenever the slit kinematics was used, only slit datapoints within 3.5 arcsec$<r<$30 arcsec were included in the modelling. For the runs with the two-component density model, we first used identical values for \mlb\ and \mld\ (runs 2A and 2B) and then we let both parameters vary with respect to each other (run 2B*). For run 2A, there were 14 values of \mbh\ ($5\times10^8$\msun\ to $3.6\times10^9$\msun) and 20 \mlb\ (3 to 13). In run 2B, we calculated models for the same 14 values of \mbh\ as in run 2A and 12 \mlb\ (4 to 10). A set of 14 \mld\ (2 to 11) values were then added for run 2B*. Run 1A and 1B used 20 \mbh\ ($5\times10^8$ to $5\times10^9$\msun) and 20 \ml\ (1 to 20).

\begin{table*}
\centering
\caption{The best-fitting \mbh, marginalised over all mass-to-light ratios, for the four quadrants and their average for all runs that we performed. All values are stated in units of $10^9$\msun. The 1\sig\ errors ($\sim68\%$ confidence level), derived from the \dchi\ analysis, are given for each quadrant. The last row gives the average of the measurements of the four quadrants with the average of their 1-\sig\ errors. We compare these errors with the parenthesised values next to them which are the standard deviation (rms) of the best-fitting values of the four quadrants.  We expect them to be comparable (see Section \ref{results}). The naming of the runs is chosen as follows. The letters A and B identify the datasets used for the kinematics. Run A used SINFONI data only while run B also used slit data within $3.5^{\prime\prime}<r<30^{\prime\prime}$. The numbers 1 and 2 indicate the use of the one-component and two-component density models respectively. In run 2A and 2B we set \mlb=\mld, whereas in run 2B* we decoupled \mlb\ and \mld. All runs adopted an inclination angle of $90^{\circ}$.}
\label{tablembh}
\begin{tabular}{llllll}
\hline
           & Run 1A                  & Run 1B                 & Run 2A                     & Run 2B                     & Run 2B* \\
           & (single-component,      & (single-component,     & (two-component (\mlb=\mld),& (two-component (\mlb=\mld),& (two-component (\mlb$\neq$\mld), \\
           & SINFONI)                & SINFONI+slit)          & SINFONI)                   & SINFONI+slit)              & SINFONI+slit) \\
\hline
Quadrant 1 & $0.97^{+0.54}_{-0.06}$   & $1.45^{+0.19}_{-0.13}$     & $1.21^{+0.31}_{-0.12}$     & $1.68^{+0.08}_{-0.21}$      & $1.68^{+0.09}_{-0.26}$ \\
Quadrant 2 & $1.21^{+0.05}_{-0.41}$   & $0.97^{+0.19}_{-0.14}$     & $1.21^{+0.23}_{-0.13}$     & $1.21^{+0.28}_{-0.08}$      & $1.21^{+0.23}_{-0.08}$ \\
Quadrant 3 & $0.97^{+0.34}_{-0.08}$   & $1.21^{+0.21}_{-0.05}$     & $1.45^{+0.07}_{-0.28}$     & $1.45^{+0.17}_{-0.07}$      & $1.45^{+0.17}_{-0.09}$ \\
Quadrant 4 & $1.45^{+0.07}_{-0.35}$   & $1.21^{+0.26}_{-0.04}$     & $1.45^{+0.12}_{-0.15}$     & $1.45^{+0.17}_{-0.09}$      & $1.45^{+0.21}_{-0.08}$ \\
\hline

Mean       &$1.15^{+0.25}_{-0.23}(0.23)$&$1.21^{+0.21}_{-0.09}(0.20)$&$1.33^{+0.18}_{-0.17}(0.14)$&$1.45^{+0.18}_{-0.11}(0.20)$&$1.45^{+0.18}_{-0.13}(0.20)$\\             
\hline
\end{tabular}
\end{table*}

\begin{table*}
\centering
\caption{The best-fitting \ml\ or \mlb\ in the $R$-band, marginalised over all \mbh, for the four quadrants and their average for all runs listed in Table \ref{tablembh}. The 1\sig\ errors ($\sim68\%$ confidence level), derived from the \dchi\ analysis, are given for each quadrant. The last row gives the average of the measurements of the four quadrants with the average of their 1\sig\ errors. Descriptions of the runs are given in the caption of Table \ref{tablembh}. For run 2B* where \mlb\ and \mld\ were decoupled, we write down only \mlb\ since \mld\ was poorly constrained. The four measurements of \mld\ in run 2B* fell within a range of 5.0 to 9.0 with an average of 8.0 and rms of 2.0} 
\label{tablemlb}
\begin{tabular}{llllll}
\hline
           & Run 1A (\ml)           & Run 1B (\ml)            & Run 2A (\mlb=\mld)        & Run 2B (\mlb=\mld)        & Run 2B*  (\mlb)\\
\hline
Quadrant 1 & $9.58^{+0.30}_{-0.99}$   & $6.74^{+0.78}_{-0.12}$     & $8.26^{+0.30}_{-1.20}$     & $6.68^{+0.35}_{-0.18}$      & $6.68^{+0.38}_{-0.38}$ \\
Quadrant 2 & $7.68^{+0.97}_{-0.31}$   & $7.68^{+0.05}_{-0.44}$     & $7.74^{+0.33}_{-0.77}$     & $7.21^{+0.29}_{-0.58}$      & $7.21^{+0.31}_{-0.37}$ \\
Quadrant 3 & $10.53^{+0.42}_{-1.48}$  & $7.68^{+0.15}_{-0.25}$     & $7.74^{+0.34}_{-0.66}$     & $7.21^{+0.24}_{-0.30}$      & $7.21^{+0.24}_{-0.33}$ \\
Quadrant 4 & $7.68^{+0.92}_{-0.26}$   & $7.68^{+0.12}_{-0.77}$     & $7.74^{+0.33}_{-0.77}$     & $7.21^{+0.13}_{-0.26}$      & $7.21^{+0.21}_{-0.46}$ \\
\hline
Mean       &$8.87^{+0.65}_{-0.76}(1.42)$&$7.45^{+0.28}_{-0.40}(0.47)$&$7.87^{+0.33}_{-0.85}(0.26)$&$7.08^{+0.25}_{-0.33}(0.26)$ & $7.08^{+0.29}_{-0.39}(0.26)$\\              
\hline
\end{tabular}
\end{table*}

\begin{figure*}
  \centering
  \includegraphics[scale=0.65]{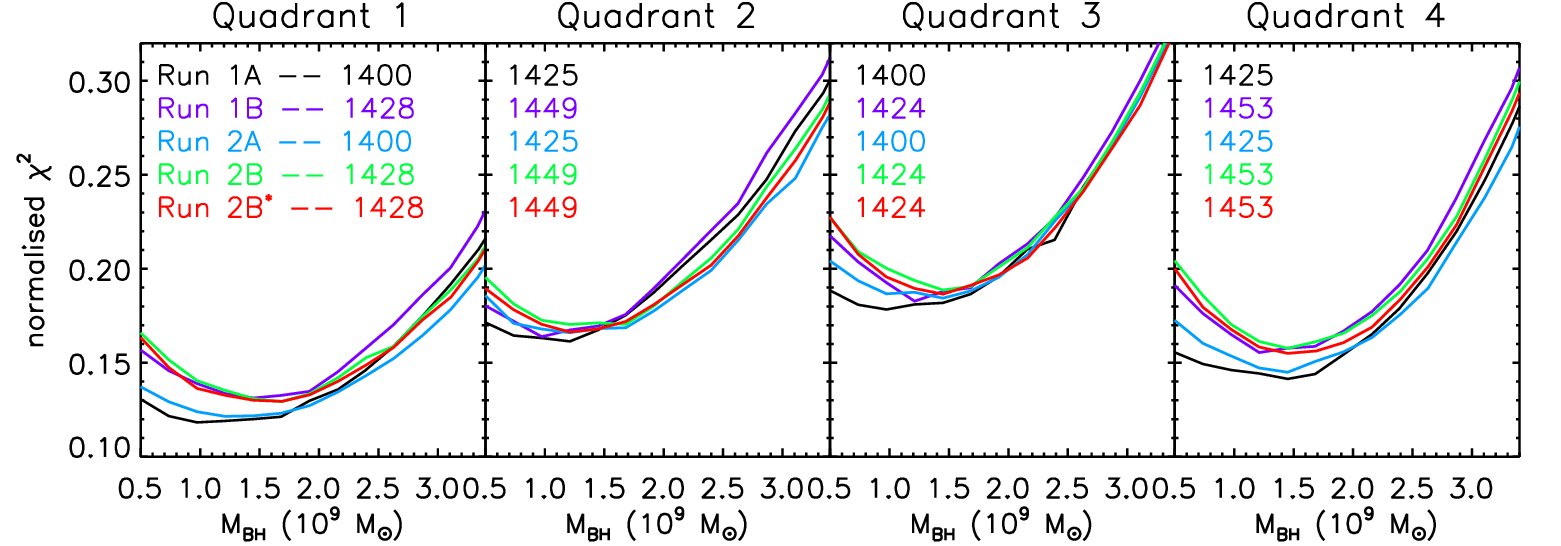}
  \caption{The normalised $\chi^2$ vs \mbh\, marginalised over all mass-to-light ratios for the runs listed in Table \ref{tablembh} for four different quadrants. The normalisation factor, by which we divide $\chi^2$, is written for the individual runs in each quadrant. The best-fitting \mbh\ is given by the model with minimum normalised $\chi^2$ in the corresponding run and quadrant. These SMBH masses can be found in Table \ref{tablembh}.}
  \label{chisqcurve}
\end{figure*}

The corresponding $\chi^2$ curve for Table 1 is shown in Fig.~\ref{chisqcurve}. We plot the normalised $\chi^2$ for all the models and all the quadrants. Ideally, the normalisation is done by dividing the $\chi^2$ by the number of degrees of freedom (dof). For the SINFONI data, the smoothing parameter included in the LOSVD derivation correlates the velocity bins, such that the number of dof is smaller than the number of observables (more details can be found in \citealt{Gebhardt-00}). Because the exact number of dof is unkown, we approximate it by the number of observables when normalising the chisquare for runs 1A and 2A. For runs 1B, 2B and 2B*, there are an additional dof contributed by the slit data. Each of the slit LOSVDs was generated from four Gauss-Hermite parameters (see Section \ref{longslitkinematics}) so there are four dof for every slit LOSVD. The normalisation factor for each run and quadrant is included in Fig.~\ref{chisqcurve}. The normalised $\chi^2$ values are less than unity, as is commonly found in the orbit-based dynamical modelling. This is partly because the number of observables is larger than the effective number of dof due to the aforementioned smoothing parameter.

For each marginalised $\chi^2$ curve of \mbh\ (see Fig. \ref{chisqcurve}) or \ml\, we approximated the 1\sig\ error (a \dchi\ of 1) by a polynomial fitting. To check whether the measurements in the four different quadrants are in agreement with each other, we compare the 1\sig\ errors with the standard deviation/rms derived from the four measurements. We expect the rms to be comparable to the 1\sig\ errors as the latter indicate the range of values within which the measurements (from quadrant to quadrant) would fluctuate. For most of the results in Table 1 and 2, the rms do not fall far from the 1\sig\ errors, especially for \mbh\ measurements. It follows that (1) the four quadrants give consistent results and (2) the $\chi^2$ analysis provides a reliable measurement of the errors in each quadrant.

All models strongly suggest the presence of a central black hole with a mass of at least $10^9$\msun. No-SMBH models were not included in any of the runs as for these models, \dchi$>20$ (compared to the best-fit model) can be readily inferred from the $\chi^2$ curves. The signature of the SMBH is strongest in the models that include long-slit data, i.e. they give higher best-fit \mbh\ in most cases and exclude the \mbh=$5\times 10^{8}$ \msun\ with higher confidence (see Fig.~\ref{chisqcurve}). \mbh=$5\times 10^8$\msun\ is the lowest \mbh\ that we modelled and it coincides with the resulting \mbh\ derived from X-ray data by H09. It is, however, not favoured by any of our runs. The $\Delta\chi^2$ of the best-fitting model for \mbh=$5 \times 10^8$\msun\ is larger than 10 for run 1A and larger than 20 for the other runs.

From the result of the five runs, we see that \mbh\ is rather sensitive to the change of the photometry and therefore the prescribed density profiles, as is also found by \citet{Siopis-09}. As the galaxy becomes rounder in the inner part of the two-component model, \mbh\ becomes systematically higher and \ml\ becomes lower compared to the one-component model. The mass-to-light ratio difference between runs 1A and 2A is larger than the difference between runs 1B and 2B. The addition of slit data reduces the difference in mass-to-light ratio but the trend remains. The increase of \mbh\ in two-component models is probably related to the flatter slope of the density inside $\sim0.3$ arcsec. It is, however, reassuring to see that the differences in the two mass models do not push \mbh\ beyond their 1\sig\ errors.

The addition of the slit data seems to bring \mlb\ to a lower value which ultimately increases \mbh\ to preserve the enclosed mass. Another important aspect from including the slit data is  that it should reduce the statistical uncertainties in the modelling. This is indeed the case as can be seen in Table 1 and 2: 1\sig\ errors decrease from run 1A to 1B and from run 2A to 2B. Fig.~\ref{chisqcurve} provides a more straightforward way to assess this. The $\chi^2$ curves of the run 1B are narrower and is enveloped by run 1A. The same is true for run 2A with respect to run 2B and also run 2B*.

In run 2B*, we repeated run 2B but allow for \mlb\ and \mld\ to be different from each other. The resulting \mbh\ and \mlb\ in all quadrants are unchanged with respect to run 2B. The best-fitting models of the four quadrants in run 2B* produce generally better $\chi^2$ values since they were given extra freedom to fit the data. It appears that untying \mld\ and \mlb\ does not lead to any change (or small if anything at all) in \mbh. \mld\ in run 2B* is not constrained well by the data since the kinematics data extend only up to 25-30 arcsec, approximately where the disc becomes important. The results suggest that the disc has a higher mass-to-light ratio than the bulge (see the caption of Table \ref{tablemlb}). Although we varied \mld\ quite extensively, \mlb\ is stable, showing that \mld\ is not correlated with \mlb.

In the dynamical modelling, the resulting mass-to-light ratio is closely connected to \mbh. It is therefore important to constrain the orbital structure in the outer part as much as possible; this gives a stronger preference for the runs which include slit data. The resulting \mbh\ of the three runs with slit data (1B, 2B and 2B*) are all consistent with each other within their 1\sig\ errors. As for the density model choice, we are more inclined towards the two-component density model. The bulge-disc decomposition was done based on the morphology of the galaxy and to allow for different components to have stellar populations with potentially different mass-to-light ratios. The results of run 2B* hint that \mld\ is higher than \mlb, although the former is not well-constrained by the data. It turns out that even though \mlb\ and \mld\ might be different, \mbh\ does not depend on a possible \ml\ gradient outside the bulge. Nevertheless, because in run 2B* we have explored the most degrees of freedom we consider its \mbh\ as the least biased estimate compared to the other runs. We selected the best-fitting model in run 2B* as our preferred model and quote the mean of \mbh\ and \mlb\ estimates in four quadrants as our best estimate. To be conservative, we adopt the largest nominal of errors as the final error margin, i.e. 0.20 for the \mbh\ and 0.39 for the \mlb.

\begin{figure*}
 \centering
 \includegraphics[scale=0.58]{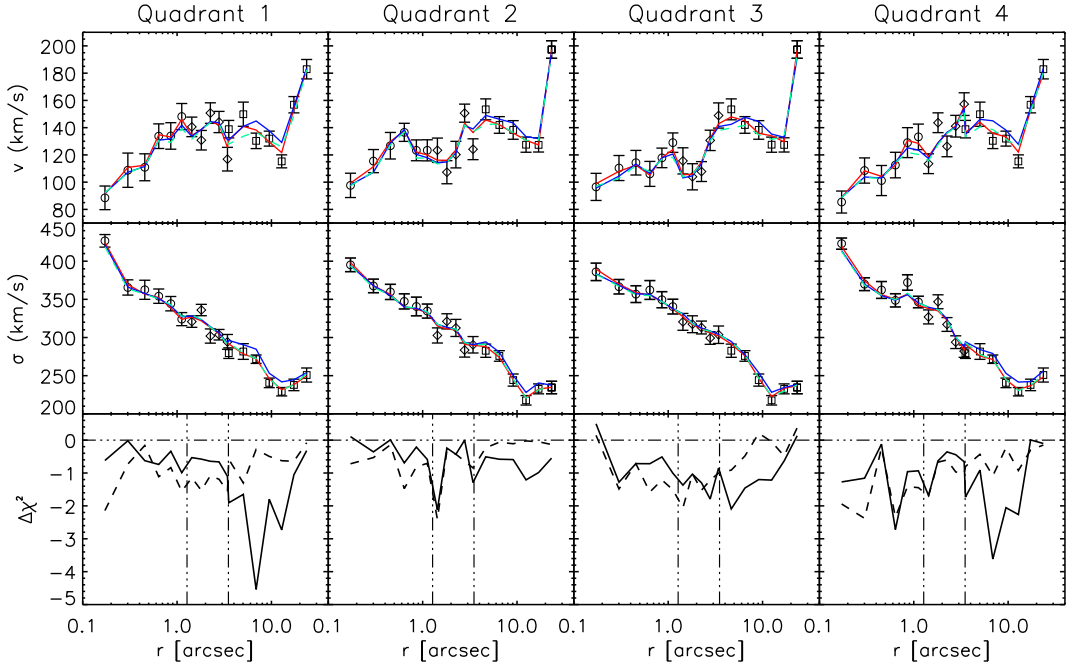}
 \caption{{\bf Top and middle rows}: kinematics fit of the models to the data along the major axis, illustrated in terms of $v$ and \sig. Circles, diamonds and squares represent the SINFONI 100mas, 250mas and the long slit data, respectively, with their corresponding errors. Our preferred model from run 2B* is shown by the red line. The blue line shows the best-fitting (smallest $\chi^2$) model for \mbh=$5\times10^8$\msun. The green line represents the model with \mbh=$5\times10^8$\msun\ and the same \mlb, \mld\ as in our preferred model. {\bf Bottom row}: $\chi^2$ differences between the three models shown in the top two rows, averaged over all angular bins in each quadrant. Following the colour coding above, solid and dashed lines represent $\chi^2_{\rm red}-\chi^2_{\rm blue}$ and $\chi^2_{\rm red}-\chi^2_{\rm green}$ respectively. The horizontal dash-three-dotted line is plotted along \dchi=0 to guide the eye and the vertical lines separate the different datasets.}
 \label{chisq}
\end{figure*}

The major-axis kinematics and the model fitting are presented for the four quadrants in the first and second row of Fig.~\ref{chisq}. We plot $v$ and \sig\ to represent the kinematics. We stress that it is the LOSVD (derived from the SINFONI spectra and from the Gauss-Hermite parameters of the slit data) that is fitted by the model, not the velocity moments. Our preferred model is shown by the red line. For comparison, we overplot the best-fitting model for \mbh\ of $5\times10^8$\msun\ (blue line). In the centre, the blue line falls below the red line as expected from the SMBH masses of the models. It climbs over the red line at intermediate radii and lies above the red line in the outer region because \mlb\ of the best-fitting model for \mbh\ of $5\times10^8$\msun\  is higher than that of our preferred model. This effect is shown by the dashed green line which represents the models where \mbh\ is $5\times10^8$ and \mlb\ and \mld\ are the same as those of the preferred model. At a first glance, it might not be directly obvious which model fits the data best. To investigate this, we plot the $\chi^2$ differences between the preferred model and the two models with \mbh=$5\times10^8$\msun\ (averaged over the five angular bins) as a function of radius in Fig.~\ref{chisq}.

The \dchi\ between our preferred model and the best-fitting model with \mbh $=5\times10^8$\msun\ is the solid line while the \dchi\ between the preferred model and the other model is the dashed line. For both cases \dchi=$\chi^2_{\rm preferred}-\chi^2_{\rm 5e8}$. Both lines generally lie at \dchi $\lesssim 0$. It is clear that models with \mbh=$5\times10^8$\msun\ produce worse fits than the preferred model. There seems to be a trade-off and an inconsistency in the models with \mbh=$5\times10^8$\msun\ with respect to the data, i.e., the mass-to-light ratio that better fits the slit data yields a worse fit to the SINFONI data. The importance of the slit data in constraining the mass-to-light ratio is apparent where the dashed lines lie mostly above the solid lines in the region covered by the slit data. For the solid line, \dchi\ is slightly lower in this region than in the SINFONI region, giving the impression that the slit data plays a more significant role in determining the black hole mass through the mass-to-light ratio. However, the dashed lines show that when the mass-to-light ratios were held fixed between models with different \mbh, there is still a strong preference for the preferred model from the SINFONI data. Both datasets have their own contributions in the SMBH mass determination and they complement each other.

One might argue that the presence of dust in this galaxy would bias the kinematics, mass tracers and the SMBH mass estimate. We compared the kinematics extracted from the dust-extincted and the dust-free part of the galaxy and we did not find systematic differences. There is also no asymmetric pattern in the SINFONI kinematics map that can be attributed to the dust (compare Fig.~\ref{sinfoimages} and Fig.~\ref{kinematicsmap}). The dust seems to be confined to a region inside $\sim0.5$ arcsec and this was masked out in the photometric analysis for the mass models. We do not observe any large-scale dust structure in NGC 1332 that would significantly affect our results (cf. \citet{Baes-00} for a discussion of the dust effects). A minor influence would probably lower SMBH mass slightly, due to the underestimated \mlb.

\subsection{Comparison with stellar population models}
\label{comparisonwithstellarpopulationmodels}
We compared the dynamical mass-to-light ratio to that of a single stellar population (\mlssp). Estimates of stellar population properties (age and metallicity for the purpose of deriving \mlssp) of NGC 1332 can be found in \citet{Barr-07} and H09. The former measurements were based on spectra from NTT-EMMI observations while the latter made use of Lick indices provided in \citet{Trager-98}. \citet{Barr-07} quote a log age of $1.103\pm0.024$ (age of $12.7\pm0.7$ Gyr) and a metallicity of  $0.270\pm0.023$ while in H09, the age is $4^{+8.8}_{-1.4}$ Gyr with a metallicity of $0.32\pm0.3$. The huge difference in the ages (apart from the errors) result in very different values of \mlssp, as derived from the SSP models of \citet{Maraston-05}. Compared to \mlssp, our mass-to-light ratio is almost two times larger when using the age of H09 with a Salpeter IMF (\mlssp$\sim 3.6$) or about three times larger with a Kroupa IMF (\mlssp$\sim 2.3$). 

Our dynamical mass-to-light ratio is much more consistent with \mlssp\ when the galaxy age is taken to be $\sim12$ Gyr. This age is the one measured by Barr et al. and it is compatible with the upper limit of H09 measurement. In addition, Lick indices measured by \citet{Ogando-08} also imply an old age of at least 12 Gyr. The Barr et al. age results in \mlssp\ of $\sim5.0$ for a Kroupa IMF or $\sim7.9$ for a Salpeter IMF. Our dynamical mass-to-light ratio falls in between these two values, but is closer to the Salpeter one. The tendency that the dynamical \ml\ agrees better with the Salpeter-based \mlssp, rather than Kroupa, in massive early-type galaxies was previously found by \citet{Cappellari-06} and also by Thomas et al. (in preparation). Although not conclusive, this might indicate that the Salpeter IMF is a better represention of stellar populations in massive early-type galaxies (see also \citealt{Grillo-09} and \citealt{Treu-09}).

\subsection{Inclination effects}
\label{inclinationeffects}
According to its high ellipticity (reaches $\sim$0.7), the orientation of NGC 1332 must be close to edge-on. To check for any influence of the residual small uncertainty in the inclination, we ran models with $i < 90$\degree. Constraining the disc to an intrinsic flattening no less than 0.2 gives 80\degree\ as the lower limit of the inclination. Using this angle, we repeated the deprojection step and the modelling for each of the four quadrants as in run 2B*. We found \mbh=$1.45\times10^9$\msun (rms=$0.20\times10^9$\msun), \mlb=6.95 (rms=0.30) and \mld=7.0 (rms=1.83). Compared to run 2B*, \mbh\ does not change and there is only a slight decrease in both \mlb\ and \mld. The uncertainties that could arise from the inclination assumption appear to be negligible in the case of NGC 1332. Compared to the preferred model ($i=90$\degree), the best-fitting model with $i=80$\degree\ is worse by a \dchi$\sim$30-50 in every quadrant.

\subsection{Dark matter halo}
Recently, concerns have emerged that dynamical models which do not include a dark halo component underestimate the true \mbh\ as is the case in M87 (\citealt{Gebhardt-09}). The reason is that models without dark halo require a higher stellar mass-to-light ratio in the outer part to compensate for the missing dark mass. In a single-component (density) model this leads to an overestimation of the central stellar mass and correspondingly to an underestimation of \mbh. 

In this work, we do not include a dark matter halo component in any of the models. Compared to the case of M87 \citep{Gebhardt-09}, the SMBH's sphere of influence in NGC 1332 is better-resolved by our SINFONI observations. The \mlb-\mbh\ degeneracy then becomes less severe as there are kinematic constraints at several radii inside the sphere of influence. This kinematic data constrain the enclosed mass at each of those radii and help to disentangle the SMBH and the stellar contribution to the central potential. We decided that our measurements are sufficiently robust and it is not necessary to further investigate \mbh\ by including a dark halo component also for the following reasons. The kinematic data that we used in the modelling only extend out to $r \lesssim 30$ arcsec or equivalently $r \lesssim 3$ kpc. At this radius, we expect only a small fraction of the total mass to be in the dark matter halo. Empirically, the dark matter fraction found in early-type galaxies with a similar luminosity to NGC 1332 (${\rm M_B} = -20.5$, as given by HyperLeda) is at most 25\% at a radius of 3 kpc \citep{Thomas-07}. The mass decomposition for NGC 1332 in H09 suggests that the dark halo only takes $\sim$10\% of the total mass at 3 kpc. Furthermore, in our preferred model, the galaxy is decomposed into a bulge and a disc with each component having its own mass-to-light ratio. The bulge is well inside the region where the dark matter is expected to be unimportant. Its mass-to-light ratio \mlb\ is then not affected by the assumption about the halo. Neglecting the halo could bias \mld\ to be too high. However, the central stellar mass is dominated by the bulge so it does not suffer from any bias in \mld\ and therefore the black hole mass is not affected by the exclusion of the dark halo. It is obvious from Run 2B and 2B* that \mlb\ is independent of the variation in \mld.

\section[]{Summary and Discussion}
\label{discussions}
We presented SINFONI observations of the lenticular galaxy NGC 1332 in the $K$-band with the purpose to measure the mass of the SMBH. The sphere of influence of the SMBH ($\sim$0.76 arcsec) is resolved by the data (spatial resolution: FWHM $\approx0.14$ arcsec). The kinematics, derived by fitting the CO bandheads, show a moderate rotation ($v \approx 150-160$ km s$^{-1}$ at the outermost radius of SINFONI data) and a high velocity dispersion ($\sim 400$ km s$^{-1}$ in the centre). To complement our SINFONI data at larger radii, we utilised major-axis long-slit data of K96. Both datasets are consistent outside $r > 2$ arcsec; inside that radius the comparison is difficult due to the PSF difference. 

To determine the SMBH mass, we performed axisymmetric Schwarzschild modelling. The systematic uncertainties inherent in the dynamical models, which are due to assumptions on the inclination, the density profile and the assumption of axisymmetry, were investigated. For the modelling input, two density models were constructed: a one-component model and a two-component model; the latter consists of a bulge and a disc. For each density model, we performed runs with and without the long-slit data. Using both SINFONI ($r < 3.5$ arcsec) and long-slit data ($3.5<r<30$ arcsec) as the kinematic constraints, SMBH masses obtained using two different density models are consistent with each other. The inclusion of long-slit data appears to give significant constraints to the models and reduce the errors. The orientation of NGC 1332 is close to edge-on and the small uncertainties in the inclination assumption causes a negligible effect on the SMBH mass. The lack of obvious signs of triaxiality, e.g., isophotal twists or kinematics misalignments, justifies the assumption of axisymmetry. \mbh\ values measured in the four different quadrants are consistent with each other.

Our preferred model is based on the two-component density profile, includes both SINFONI and long-slit data and adopts an inclination angle of 90\degree. We find a SMBH mass of  \mbh=$(1.45\pm0.20)\times10^9M_\odot$ and a bulge mass-to-light ratio \mlb=$7.08\pm0.39$ in the $R$ band. The disc mass-to-light ratio is not well-constrained by the data but falls within a range of 5.0 to 9.0 (also in $R$-band). The comparison of our results to previous measurements and the implication of this \mbh\ for the \msig\ and the \mlum\ relations are discussed in the following.

\subsection{Comparison with results from X-ray data}
Our results were preceeded by an X-ray analysis of NGC 1332 using Chandra data (H09). They derive a most probable SMBH mass of $5.2^{+4.1}_{-2.8} \times10^8$\msun\ ($5.4^{+4.3}_{-2.9}\times10^8$\msun\ at the distance adopted in this work), and a mass-to-light ratio in $J$-band of $1.16^{+0.12}_{-0.14}$ ($1.11^{+0.11}_{-0.14}$ at our distance). As shown in Section \ref{results}, models with \mbh=$5\times10^8$\msun\ do not provide a proper fit to the kinematic data and they are separated from our preferred model by \dchi\ of at least 40. This is understandable since the SINFONI data give a high central velocity dispersion inside the sphere of influence and thus such a SMBH mass would be incommensurate. The \mbh\ value of H09 is obtained by using the \msig\ relation as the Bayesian prior. They are unable to rule out the possibility that the SMBH mass is an underestimate. The use of X-ray data alone, without the \msig\ prior, decreases their \mbh\ estimate even further although the upper limit is similar to our \mbh\ estimate (see their Fig. 8).

\begin{figure}
 \centering
 \includegraphics[scale=0.48]{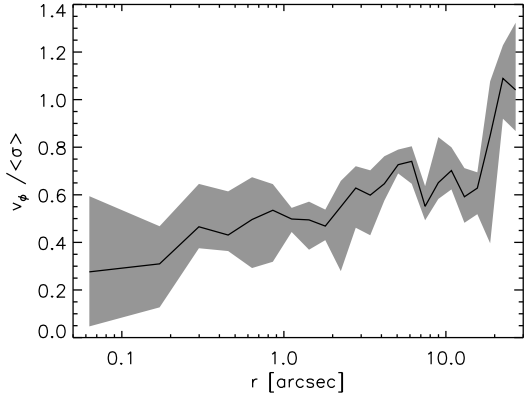}
 \caption{The significance of rotational velocity relative to the random motion of the stars as a function of radius along the major axis. $v_\phi$ is the mean rotation in the azimuthal direction where $(v_\phi^2+\sigma_\phi^2)$ constitutes the second moment of the azimuthal velocity. $\langle\sigma\rangle$ is the local mean velocity dispersion defined as ${\langle\sigma\rangle}^2=(\sigma_r^2+\sigma_\theta^2+\sigma_\phi^2)/3$. The shade shows the area where all models within 1\sig\ error fall and the line represents the average.}
 \label{rotation}
\end{figure}

Comparing our stellar mass profile with that of H09, we find that while the slope is very similar, our enclosed mass is twice as high at all radii where the stars dominate. The discrepancy is, however, not specific to H09; stellar mass profiles for NGC 1332 derived from other X-ray analyses (\citealt{Fukazawa-06}; \citealt{Nagino-09}) seem to be in agreement with H09. We find that lowering the inclination to 80\degree only has a negligible effect to the mass profile. From the orbital structure of our preferred model we find that the rotation is only important at the outer radii. In Fig.~\ref{rotation}, we plot $v_\phi/\langle\sigma\rangle$ along the major axis, where the rotation is highest. This quantity measures the importance of rotation with respect to random motion. At the inner radii, the significance of rotation is low and thus negligible for the mass estimation. The $v_\phi/\langle\sigma\rangle$ profile, however, rises with radius and peaks at a value of order unity at $r\sim20$ arcsec. From this, we can expect that if the gas follows the stellar rotation, the X-ray-derived mass would be underestimated the most around this radius by roughly 25\%. Rotation can therefore not be responsible for the systematic difference in the mass profile. The difference in mass is reflected by the difference in the resulting mass-to-light ratios in our models and the models of H09. From the stellar population point of view, the low mass-to-light ratio measured by H09 can be obtained if this galaxy is young. However, the age of 4 Gyr as derived in H09 has a large upper error which is consistent with the old age found by \citet{Barr-07} (see subsection \ref{comparisonwithstellarpopulationmodels}).

This is not the first time that a dynamically derived mass differs from that of the X-ray (\citealt{Shen-09}; \citealt{Johnson-09}; \citealt{Romanowsky-09}). In those works, the invalidity of the assumption of hydrostatic equilibrium is suspected to contribute at least to some fraction of the discrepancy, as is also pointed out by \citet{Diehl-07} and \citet{Ciotti-04}. Inflowing gas and the presence of non-thermal pressure due to magnetic fields, microturbulence or cosmic rays are possible reasons as to why the X-ray derived mass can be lower than the true mass (\citealt{Ciotti-04}; \citealt{Johnson-09}; \citealt{Churazov-08}). In the case of NGC 1332, our dynamical modelling implies a much higher \mbh, which might indicate the failure of hydrostatic approximation, at least in the inner part. In the outer region, rotation can possibly account for a small part of the missing X-ray mass. All in all, the mass discrepancy probably involves a combination of systematic uncertainties in both methods, for which a detailed inspection is outside the scope of this paper. A situation similar to that of NGC 1332 has been recently reported by \citet{Shen-09} for NGC 4649. Their orbit-based modelling results in a larger \mbh\ and a $\sim 70\%$ higher mass profile than obtained from X-rays in H09. A larger sample of galaxies would indeed be required to investigate whether this trend applies generally for these two methods. 

\subsection{SMBH-bulge relation}

\begin{figure}
  \centering
  \includegraphics[scale=0.45]{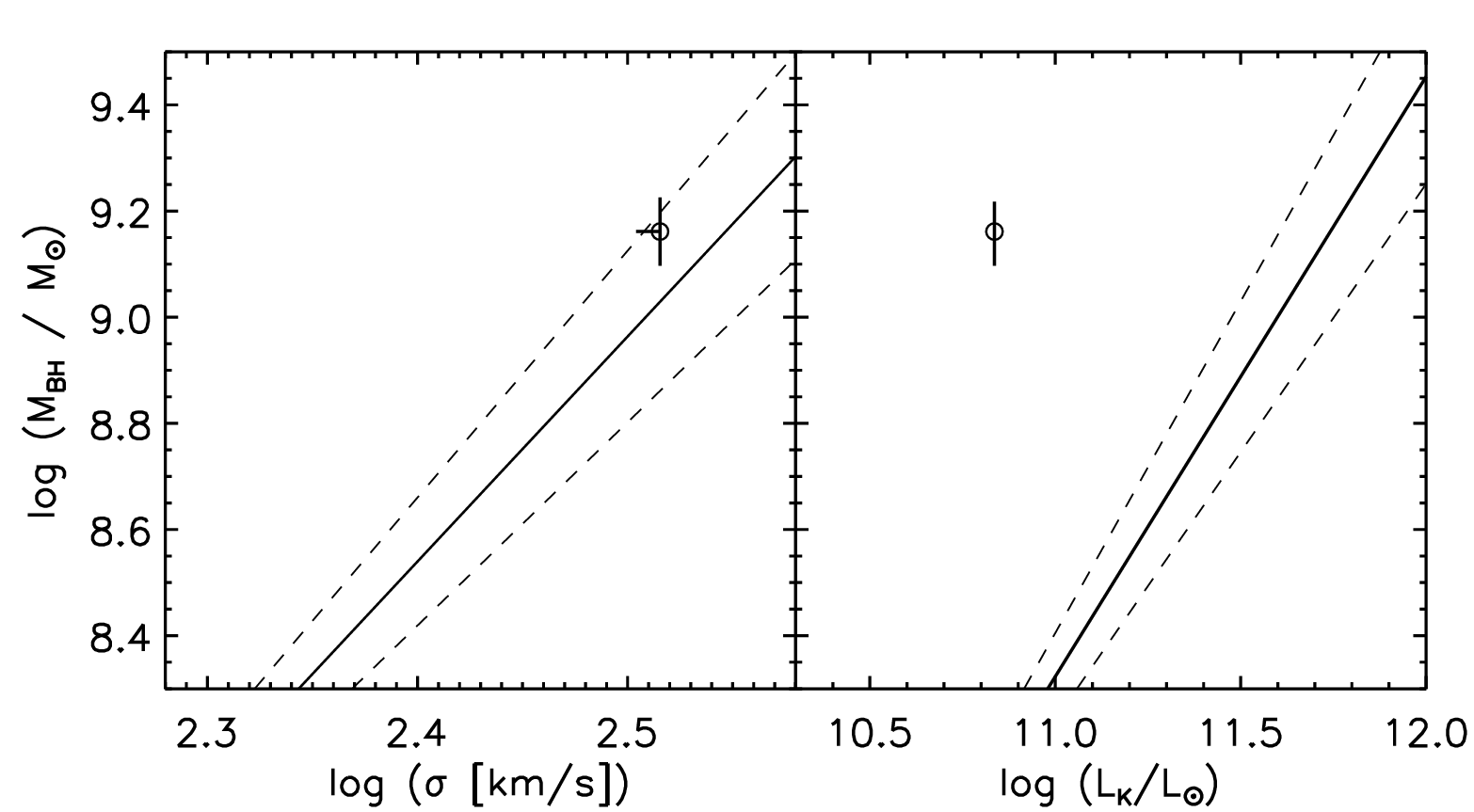}
  \caption{\msig\ (left) and \mlum\ (right) diagrams. NGC 1332 is plotted as a square in each panel along with the \citet{Gueltekin-09} \msig\ relation and Marconi\&Hunt (2003) \mbh-$L_K$ relation.}.
  \label{relations}
\end{figure}

With our \mbh\ measurement, and our photometric decomposition, NGC 1332 is displaced from the standard relations in both \msig\ and \mlum\ diagrams (Fig.~\ref{relations}). The galaxy is located above the \citet{Gueltekin-09} \msig\ relation by 0.15-0.20 dex which is still within the intrinsic scatter. The velocity dispersion of 327.7 km s$^{-1}$ was calculated using the definition of $\sigma_e$ in \citet{Gueltekin-09}. It was measured using the slit data within the effective radius of the bulge of $8.39^{\prime\prime}$. Without the luminosity-weighting, the velocity dispersion drops to 319.2 km s$^{-1}$ which is very close to the value found in HyperLeda. 

In the \mlum\ diagram, NGC 1332 is strikingly off of the \citet{Marconi-03} relation. Our \mbh\ is one order of magnitude higher than expected for its bulge luminosity. If the $M_{BH}$-$L_K$ ($L_K$ is the bulge luminosity in the $K$-band) were obeyed, it would result in a black hole mass of $1.37\times10^8$\msun\ which is highly excluded in any of our runs. 

The \sig-$L$ relation in the current black hole samples is known to be different from that in the SDSS sample, on which the distributions of $L$ and \sig\ are based; the black hole samples have larger \sig\ for a given $L$ or smaller $L$ for a given \sig\ (\citealt{Bernardi-07}; \citealt{Tundo-07}; \citealt{Lauer-07}). In this case NGC 1332 is not an exception. It is in fact a rather extreme outlier in the \sig-$L$ diagram of \citet{Bernardi-07} for SMBH sample, i.e. the bias is stronger than expected from the SMBH sample. The \msig\ and the \mlum\ relations predict different SMBH masses and contradict each other by a factor of about seven.
 
Provided that the bias in the \sig-$L$ relation is just a selection effect, a question arises: which relation is the more fundamental one? Our result for NGC 1332 favours the \msig\ to be the more fundamental relation since the measured \mbh\ for this galaxy turns out to fall much closer to the value predicted by the \msig\ relation. This is in line with the suggestion of \citet{Bernardi-07}. In addition, NGC 1332 is located slightly above the \msig\ relation, which makes it also consistent with the suggestion that the relation curves upwards at the upper end (\citealt{Wyithe-06}) or that the intrinsic scatter increases in this regime.

\section*{Acknowledgements}
We thank the Paranal Observatory Team for support during the observations. We are grateful to Koenraad Kuijken for sharing the long slit spectral data which we used for the dynamical modelling and to Ortwin Gerhard for useful discussions. We would also like to acknowledge the anonymous referee, whose comments have improved this paper. The research of P.E. is supported by the Deutsche Forschungsgemeinschaft through the Priority Programme 1177 'Galaxy Evolution'. Support for N.N. is provided by the Cluster of Excellence: 'Origin and Structure of the Universe'.

\bibliographystyle{mn2e}
\bibliography{bibliography}

\appendix

\section[]{Seeing correction in the deprojection}
\label{appendix}
Seeing due to atmospheric turbulence flattens the slope of the light profile of extended objects obtained by ground-based observations, especially at the innermost radii. When the surface brightness profile is deprojected without PSF correction, the resulting luminosity density will also be flatter than it actually is. Since light traces stars this will then lower the stellar mass contribution. The dynamical modelling only constrains the total mass, and so \mbh\ will be overestimated to compensate for the decreased mass in stars. A priori, it is not known how much this effect alters the SMBH mass.  

We estimated the PSF effect using two one-component models described in Section \ref{photometricmodels}. The first was the one-component model used in the main analysis, where we relied on the SINFONI photometry (i.e., the collapsed datacube) for the innermost isophotes (hereafter referred to as the SINFONI dataset). The second one was the same model but we instead used the isophotes derived from the WFPC2 image to replace the SINFONI photometry at $r<0.5$ arcsec (HST dataset). We accounted for the seeing in the deprojection step described in Section \ref{deprojection}. 

\begin{figure}
  \centering
  \includegraphics[scale=0.48]{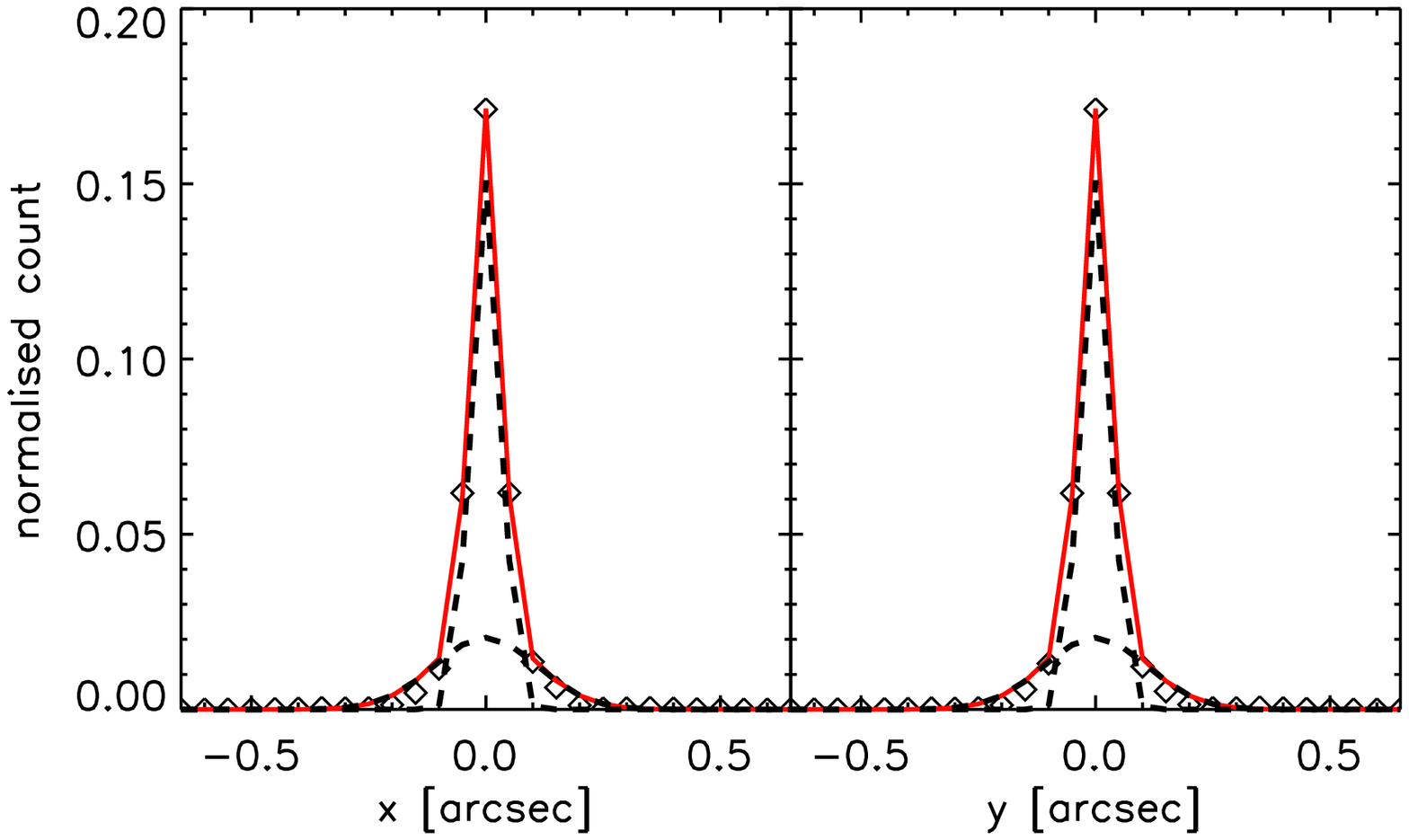}
  \caption{A double non-circular Gaussian fit to the HST PSF. The fit along the x and the y-axis of the detector are shown in the left and right panel respectively. The black dashed lines are the individual Gaussians and the red solid lines are the sum of the Gaussian components; diamonds are the actual PSF.}
  \label{hstpsf}
\end{figure}

For the HST dataset, we generated a PSF for the Planetary Camera chip of WFPC2 with version 6.3 of the TinyTim software package,\footnote{http://www.stsci.edu/software/tinytim/tinytim.html} using the location of the galaxy center and an appropriate K giant spectrum. This instrumental PSF of the WFPC2 image was then parametrised in the same way as that of the SINFONI image, i.e. as a non-circular double Gaussian function. Fig.~\ref{hstpsf} shows the fit to the HST PSF. The resulting FWHMs are 0.067 arcsec and 0.24 arcsec for the narrow and broad components respectively (${\rm FWHM_x}\approx {\rm FWHM_y}$ for both components). 

\begin{figure}
\centering
 \includegraphics[scale=0.48]{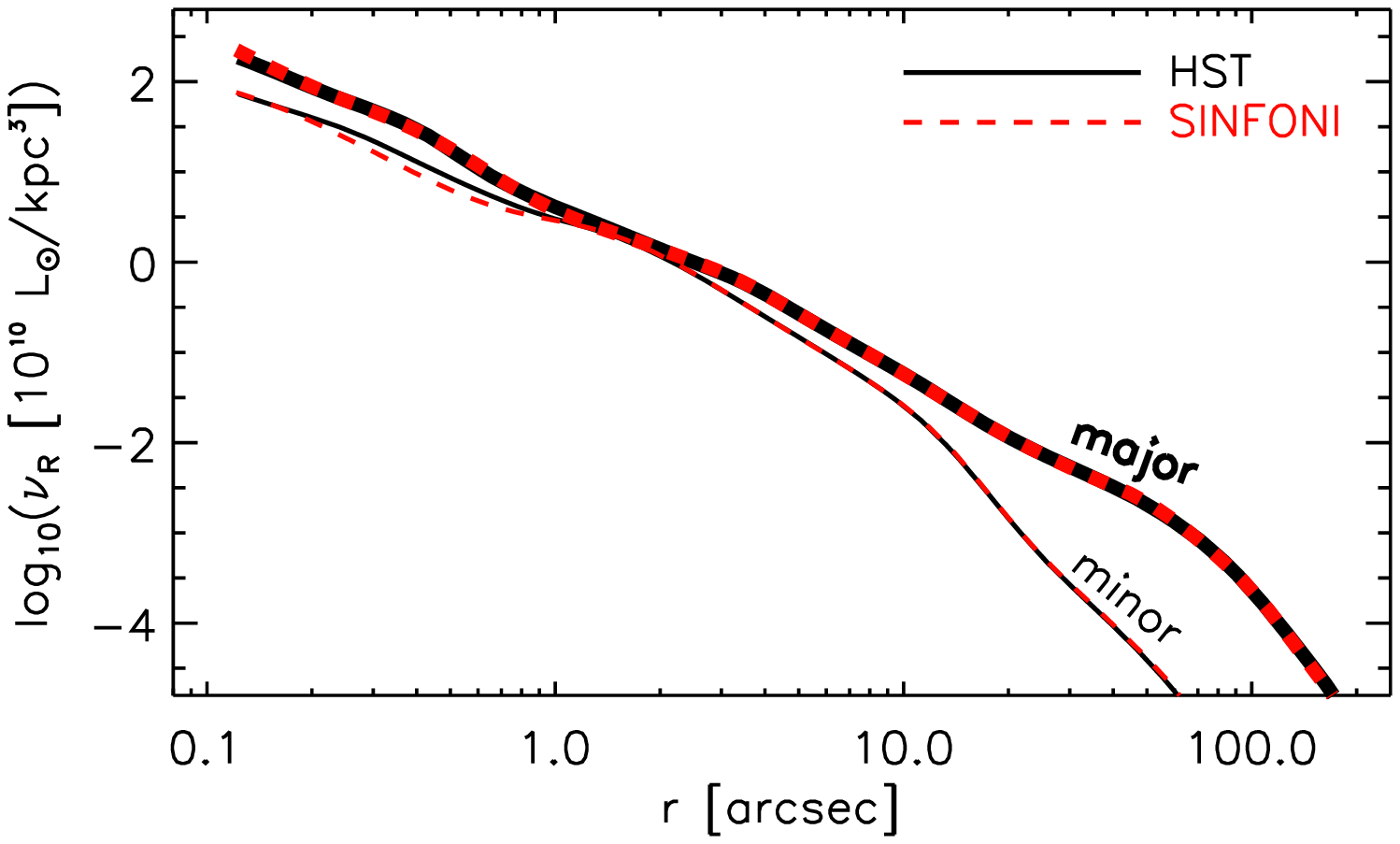}
 \includegraphics[scale=0.48]{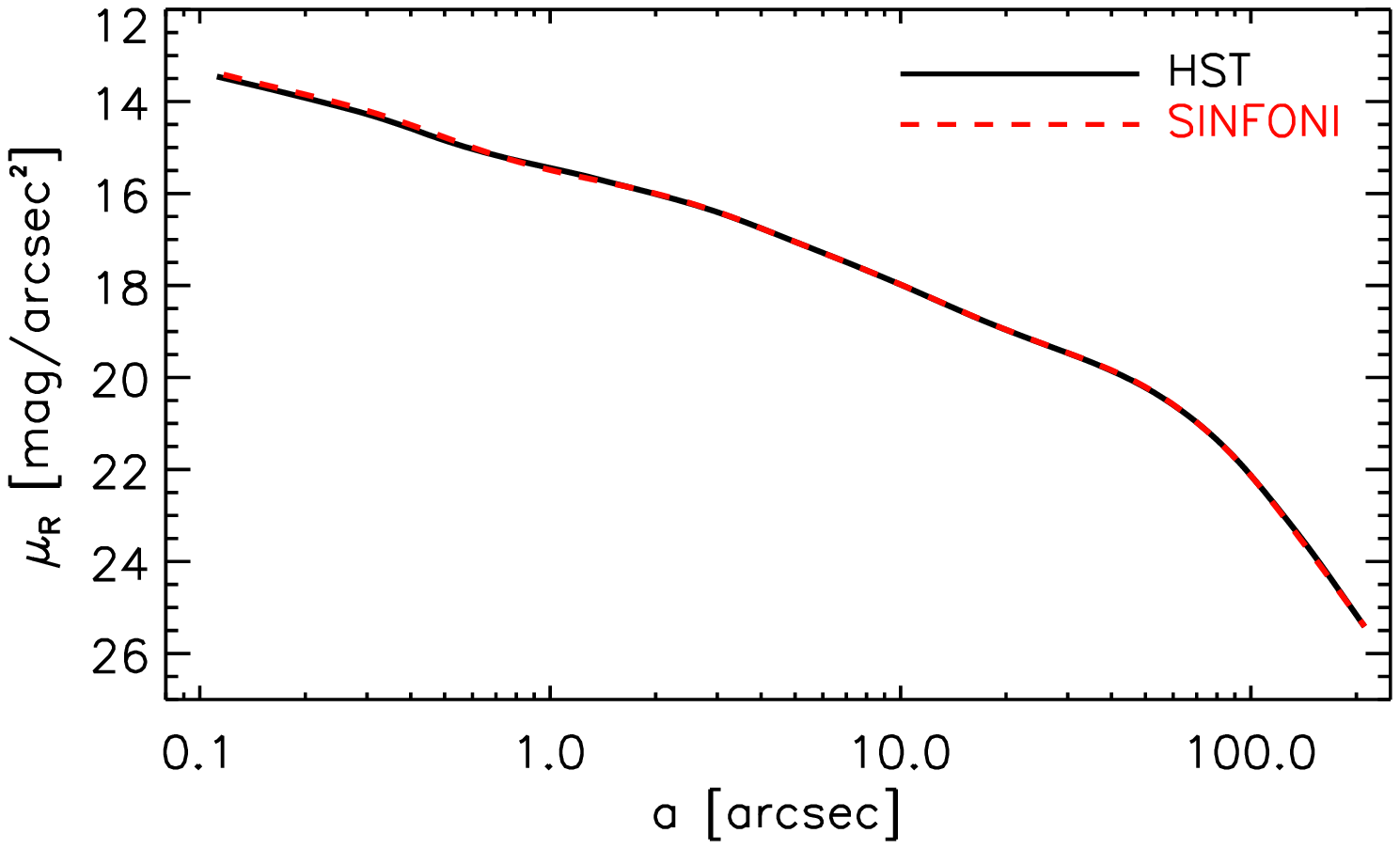}
 \caption{The PSF-deconvolved luminosity density models (top panel). Thick and thin lines refer to density profiles along the major and minor axes respectively. The reprojection of these luminosity density models without seeing convolution resulted in the PSF-deconvolved surface brightness models (bottom panel). The SINFONI dataset (red dashed line) used SINFONI images out to $4.5$ arcsec and the HST dataset (black solid line) used HST images to replace SINFONI data for the innermost isophotes ($r<0.5$ arcsec). For $r>4.5$ arcsec, the NTT-EMMI image was used for both profiles.}
\label{lumdens2}
\end{figure}

The top panel of Fig.~\ref{lumdens2} shows the luminosity density profiles along the major and minor axes for both HST and SINFONI datasets. Both densities are almost identical at all radii along the major axis. There is only a small difference along the minor axis at 0.2 arcsec\,$\lesssim r \lesssim$0.8 arcsec, implying that the SINFONI density model is slightly more flattened at those radii. Due to the strong dust extinction in the WFPC2 image, the ellipticity in the inner part is not well constrained and is fixed to an approximate value. Together with the different shape of the HST and the SINFONI PSF, this prevents an exact match of SINFONI and HST density profiles. However, the general agreement between the two profiles is very good considering that the inner profile was constructed using different datasets, which were deconvolved using different PSF images.

The bottom panel shows the reprojection of the luminosity density models of both datasets without seeing convolution, which should reflect the intrinsic surface brightness profiles. We see that both profiles agree well with each other in the inner part where the correction is most significant ($r\lesssim 1$ arcsec). In the outer part, the profiles overlap with each other, as is also the case in the luminosity density profiles. This is expected as both profiles use the same EMMI data at $r > 4.5$ arcsec and the effect of seeing is restricted only to the innermost isophotes.

We performed the dynamical modelling using each of the luminosity density models from the two datasets, once with PSF correction and once without. The modelling setups were identical to those in Section \ref{dynamicalmodelling}. Our parameter grid for each of the runs consisted of 20 trial values of \mbh\ ranging from $5\times10^8$\msun\ to $5\times10^9$\msun, each was paired with 20 different \ml\ values ranging from 1 to 10. To minimise computing time, we only used SINFONI kinematics. \mbh\ is sensitive to the change of the density profile and using only SINFONI data will put the least constraints to \mbh. Therefore in some sense these runs should show the largest possible change in \mbh\ due to the PSF inclusion.  

We list the best-fitting \mbh\ (average of four quadrants), marginalised over \ml, together with the averaged 1\sig\ errors in Table \ref{tableappendix}. For comparison, we rewrite the results of run 1A in Table \ref{tablembh} as the model ``SINFONI with PSF correction''. When the seeing effect on the photometry is taken into account, \mbh\ decreases. This is expected, as the PSF deconvolution steepens the slope of the surface brightness and the luminosity density profile, giving more mass to the stars. The mass-to-light ratio \ml\ increases only slightly, from $\sim8.6$ for models without PSF correction to $\sim8.9$ when either of the PSFs is included. The photometric difference (between the HST and the SINFONI image) in the innermost region does not seem to affect \ml.

\begin{table}
\centering
\caption{SMBH masses obtained for different modelling runs with different luminosity models. The given errors are the average of 1\sig\ errors from the four quadrants. All \mbh\ are given in units of $10^9$\msun.}
\label{tableappendix}
\begin{tabular}{lll}
\hline
Luminosity model               & \mbh            \\
\hline
HST without PSF correction     & $1.56\pm0.24$   \\
SINFONI without PSF correction & $1.92\pm0.24$   \\
HST with PSF correction        & $1.27\pm0.22$   \\
SINFONI with PSF correction    & $1.15\pm0.24$   \\
\hline
\end{tabular}
\end{table}

For the runs where the PSF is not included, the SINFONI \mbh\ is higher than that of HST. When the PSF is included, the SINFONI \mbh\ decreases more dramatically (see Table \ref{tableappendix}). These changes are expected, as the HST PSF is narrower than the SINFONI PSF. When the HST data are used, the change of \mbh\ due to the PSF is still within the 1\sig\ error, while for SINFONI data the change is larger, but still less than a factor of two. Although this effect would not apply equally to all galaxies and observation modes, our exercise seems to suggest that the PSF can be considered a relatively minor issue for HST photometry. However, for ground-based observations which in general have broader PSFs, more care is needed. 
 
Comparing \mbh\ obtained using the HST and the SINFONI datasets after PSF deconvolution, we see that both masses are consistent and lie well within their 1\sig\ errors. This result is reassuring considering the uncertainties in the SINFONI PSF.

\end{document}